\documentclass[vecphys]{svmult}

\usepackage{makeidx}         
\usepackage{graphicx}        
\usepackage{multicol}        
\usepackage[bottom]{footmisc}

\usepackage{amsmath}
\usepackage{amssymb}

\newcommand{\TeV}{\,\mathrm{TeV}}

\newcommand{\MeV}{\,\mathrm{MeV}}
\newcommand{\ie}{i.e.~}

\newcommand{\diag}{\negmedspace \mathrm{diag}}

\newcommand{\uPl}{\mathrm{\sss{Pl}}}
\newcommand{\mpl}{m_{_{\uPl}}}
\newcommand{\sss}[1]{{\scriptscriptstyle{#1}}}

\newcommand{\zero}{{\sss{0}}}
\newcommand{\one}{{\sss{1}}}
\newcommand{\two}{{\sss{2}}}

\newcommand{\calH}{\mathcal{H}}

\newcommand{\calF}{\mathcal{F}}

\newcommand{\calS}{\mathcal{S}}

\newcommand{\calM}{\mathcal{M}}

\newcommand{\calP}{\mathcal{P}}
\newcommand{\ueq}{\mathrm{eq}}
\newcommand{\urad}{\mathrm{rad}}

\newcommand{\uini}{\mathrm{ini}}
\newcommand{\uic}{\mathrm{ic}}
\newcommand{\ure}{\mathrm{re}}
\newcommand{\uquant}{\mathrm{q}}

\newcommand{\uend}{\mathrm{end}}
\newcommand{\unuc}{\mathrm{nuc}}

\newcommand{\udm}{\mathrm{c}}
\newcommand{\uM}{\mathrm{M}}
\newcommand{\ureh}{\mathrm{reh}}

\newcommand{\ud}{\mathrm{d}}
\newcommand{\ue}{\mathrm{e}}

\newcommand{\ub}{\mathrm{b}}

\newcommand{\uS}{\sss{\mathrm{S}}}
\newcommand{\uT}{\sss{\mathrm{T}}}

\newcommand{\mukha}{Q}
\newcommand{\qmode}{\mu}
\newcommand{\qmodeS}{\qmode_\uS}
\newcommand{\qmodeT}{\qmode_\uT}
\newcommand{\source}{\calS}
\newcommand{\tensor}{h}
\newcommand{\moduli}{\chi}
\newcommand{\mat}{\varphi}

\newcommand{\Vall}{V}
\newcommand{\Vgrav}{W}
\newcommand{\Vmat}{U}

\newcommand{\const}{C}
\newcommand{\constdec}{\const_\uquant}

\newcommand{\Afac}{A}

\newcommand{\epsone}{\epsilon_\one}
\newcommand{\epstwo}{\epsilon_\two}

\newcommand{\hubble}{H}

\newcommand{\kpivot}{\kwav_*}

\newcommand{\firstform}{\eta}
\newcommand{\unitv}{u}
\newcommand{\kwav}{k}

\newcommand{\adia}{\sigma}

\newcommand{\metric}{\ell}
\newcommand{\christoffel}{\Upsilon}
\newcommand{\field}{\calF}

\newcommand{\Phipert}{\Phi}
\newcommand{\Psipert}{\Psi}

\newcommand{\fieldpert}{\delta \field}
\newcommand{\adiapert}{\delta \adia}
\newcommand{\curvpert}{\zeta}
\newcommand{\entropert}{\delta \varsigma}
\newcommand{\orthometric}{\perp}

\newcommand{\power}{\calP}
\newcommand{\obspert}{\nu}

\newcommand{\powerTens}{\power_\tensor}
\newcommand{\OmegaCDM}{\Omega_\udm}

\newcommand{\OmegaL}{{\Omega_{\Lambda}}}
\newcommand{\OmegaR}{{\Omega_{\urad}}}
\newcommand{\Mpc}{\mathrm{Mpc}}
\newcommand{\camb}{\textsc{camb}}
\newcommand{\cosmomc}{\textsc{cosmomc}}

\newcommand{\OmegaB}{\Omega_\ub}

\newcommand{\optdepth}{\tau}

\newcommand{\reheat}{R}
\newcommand{\lnR}{\ln \reheat}
\newcommand{\usssPMoneP}{\sss{(\uM=1)}}
\newcommand{\calPnum}{\calP^{\usssPMoneP}}
\newcommand{\wstate}{w}

\makeindex             

\begin{document}

\title*{The exact numerical treatment of inflationary models}
\author{Christophe Ringeval}
\institute{Theoretical and Mathematical Physics Group, Centre for
  Particle Physics and Phenomenology, Louvain University, 2 Chemin du
  Cyclotron, 1348 Louvain-la-Neuve, Belgium.}
%
%

\maketitle

\begin{abstract}

The precision reached by the recent CMB measurements gives new
insights into the shape of the primordial power spectra of the
cosmological perturbations. In the context of inflationary cosmology,
this implies that the CMB data are now sensitive to the form of the
inflaton potential. Most of the current approaches devoted to the
derivation of the inflationary primordial power spectra, or to the
inflaton potential reconstruction problem, rely on approximate
analytical treatments that may break down for exotic models. In this
article, we numerically solve the inflationary evolution of both the
background and all the perturbed quantities to extract the primordial
power spectra exactly. Such a method solely relies on General
Relativity and linear perturbation theory. More than providing a tool
to test analytical approximations, one may consider, without
complications, the treatment of non-standard inflationary models as
those involving several fields, eventually non-minimally coupled to
gravity.

The usefulness of the exact numerical approach to deal with CMB data
is illustrated by analysing the WMAP third year data in the context of
single field models. For this purpose, we introduce a new inflationary
related parameter encoding the basic properties of the reheating
era. This reheating parameter has significant observable effects and
provides a self-consistency test of inflationary models. As a working
example, the marginalised probability distributions of the reheating
and potential parameters associated with the small field models are
presented.

\end{abstract}

\section{Motivations}
\label{sect:intro}

The inflationary paradigm is currently passing all the tests raised by
the so-called high precision cosmology
measurements~\cite{Spergel:2006hy}. Although this suggests that the
existence of a quasi-exponential accelerated era in the early universe
may be viewed as a standard lore, one has to keep in mind that almost
all the inflationary field models lasting more than sixty efolds and
leading to an almost scale-invariant power spectrum for adiabatic
scalar perturbations may do the job. It is therefore of both
theoretical and observational interest to look for inflationary
properties that are, or will be in a foreseeable future, significant
enough in the data to allow disambiguation between the different
models. Many works are devoted to this task ranging from the details
of the reheating era to the search of a theoretical embedding of
inflation in supersymmetry or string theory~\cite{Bassett:2005xm,
Allahverdi:2006iq, Allahverdi:2006we, Tye:2006uv}. In the following,
we will be interested in the model disambiguation problem through the
cosmological perturbations and the Cosmic Microwave Background (CMB)
anisotropies.

Among the analytical tools available to study inflation in the
cosmological context, the so-called slow-roll approximation provides
analytical expressions for both the field evolution and the primordial
scalar and tensor power spectra. It relies on an order by order
expansion in terms of the so-called Hubble-flow functions
$\epsilon_i(n)$, where $n=\ln (a/a_\uini)$ is the number of efolding
from the beginning of inflation and $a$ the
Friedman-Lema{\^\i}tre-Robertson-Walker (FLRW) scale
factor~\cite{Schwarz:2001vv}. The Hubble flow fonctions are defined
from the Hubble parameter $\hubble(n)$ by
\begin{equation}
\epsone = -\dfrac{\ud \ln \hubble}{\ud n}, \qquad \epsilon_{i+1} =
\dfrac{\ud \ln \epsilon_i}{\ud n}.
\end{equation}
If the underlying field model is such that these functions remain
small at the time where the length scales of cosmological interest
today leave the Hubble radius, then the scalar and tensor power
spectra can be Taylor expanded around a given pivot wavenumber
$\kpivot$. At first order, one gets~\cite{Lyth:1998xn,
Mukhanov:1990me, Stewart:1993bc, Martin:1999wa}
\begin{equation}
\label{eq:srpowerscal}
\calP_\zeta(k) = \dfrac{\kappa^2 H^2}{8 \pi^2\epsone } \left[ 1 - 2
  \left(C+1 \right) \epsone - C \epstwo - \left(2\epsone +
  \epstwo\right) \ln \dfrac{k}{\kpivot} \right],
\end{equation}
for the scalar modes and
\begin{equation}
\label{eq:srpowertens}
\calP_\tensor(k) = \dfrac{2
  \kappa^2 H^2}{\pi^2} \left[1 - 2 \left(C+1 \right) \epsone
  -2 \epsone \ln \dfrac{k}{\kpivot}\right],
\end{equation}
for the tensor modes. In Eqs.~(\ref{eq:srpowerscal}) and
(\ref{eq:srpowertens}), $\kappa^2 = 8 \pi/\mpl^2$ is the gravitational
coupling constant and $C$ is a constant coming from the Taylor
expansion ($C\simeq -0.73$). The Hubble parameter and the two first
Hubble flow functions are evaluated at $N_* = n_\uend - n_*$: the
number of efold before the end of inflation at which the pivot length
scale crosses the Hubble radius: $\kpivot = a(N_*) \hubble(N_*)$.
\begin{figure}
\begin{center}
\includegraphics[width=10cm]{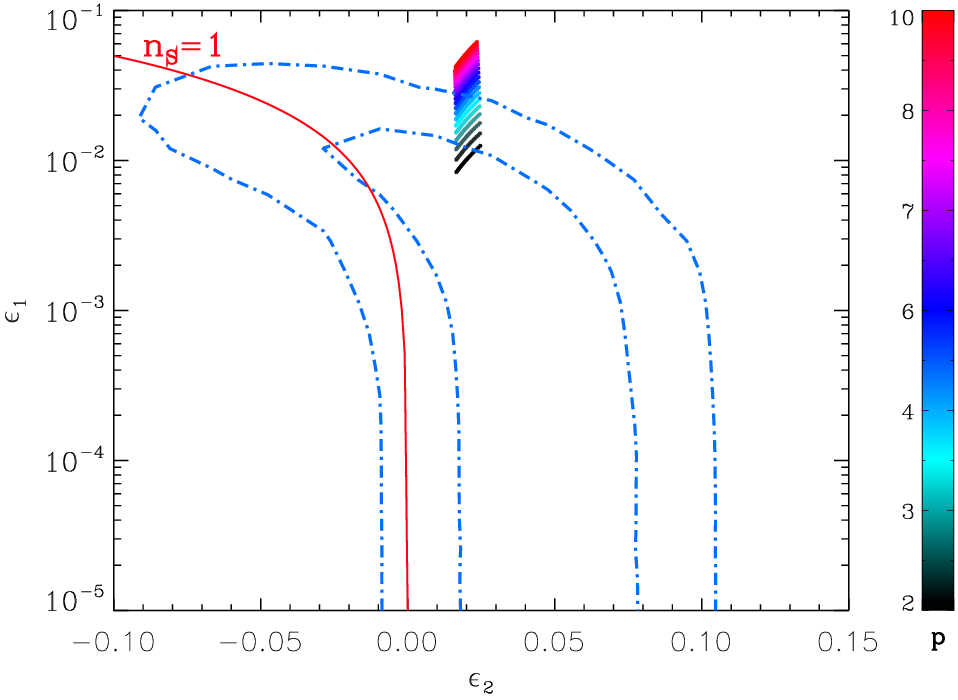}
\caption{WMAP third year data constraints on the first order Hubble
  flow parameters $\epsone(N_*)$ and
  $\epstwo(N_*)$~\cite{Martin:2006rs}. The two dashed contours are the
  $68\%$ and $95\%$ confidence intervals associated with the
  two-dimensional marginalised posterior probability distribution. The
  solid curve corresponds to a scale invariant power spectrum whereas
  the short segments are the slow-roll predictions for the large field
  models $V(\mat) \propto \mat^p$. Note that the model predictions are
  not ``dots'' in the plane $(\epsone,\epstwo)$ due to their
  dependence with respect to $N_*$. Indeed, due to uncertainties on
  the reheating era, the efold $N_*$ for which the observable pivot
  scale leaves the Hubble radius during inflation is not known (see
  Fig.~\ref{fig:scales}). However, under reasonable assumptions, one
  may assume $ 40 \lesssim N_* \lesssim 60$ thereby leading to a
  ``segment'' in the plane $(\epsone,\epstwo)$.}
\label{fig:srwmap}
\end{center}
\end{figure}
From these power spectra, assuming the conservation of the comoving
curvature perturbation after Hubble exit ($\kwav < a \hubble$), the
CMB anisotropies induced by the scalar and tensor perturbations can be
derived and compared with the data. Using Markov-Chains-Monte-Carlo
(MCMC) methods, one can extract constraints on the power spectra
parameters, namely $\epsone$, $\epstwo$ and $P_*=\kappa^2 \hubble^2/(8
\pi^2 \epsone)$. Assuming a flat FLRW universe, the WMAP third year
data lead to the posterior probability distributions plotted in
Fig.~\ref{fig:srwmap}.

\begin{figure}
\begin{center}
\includegraphics[width=11.5cm]{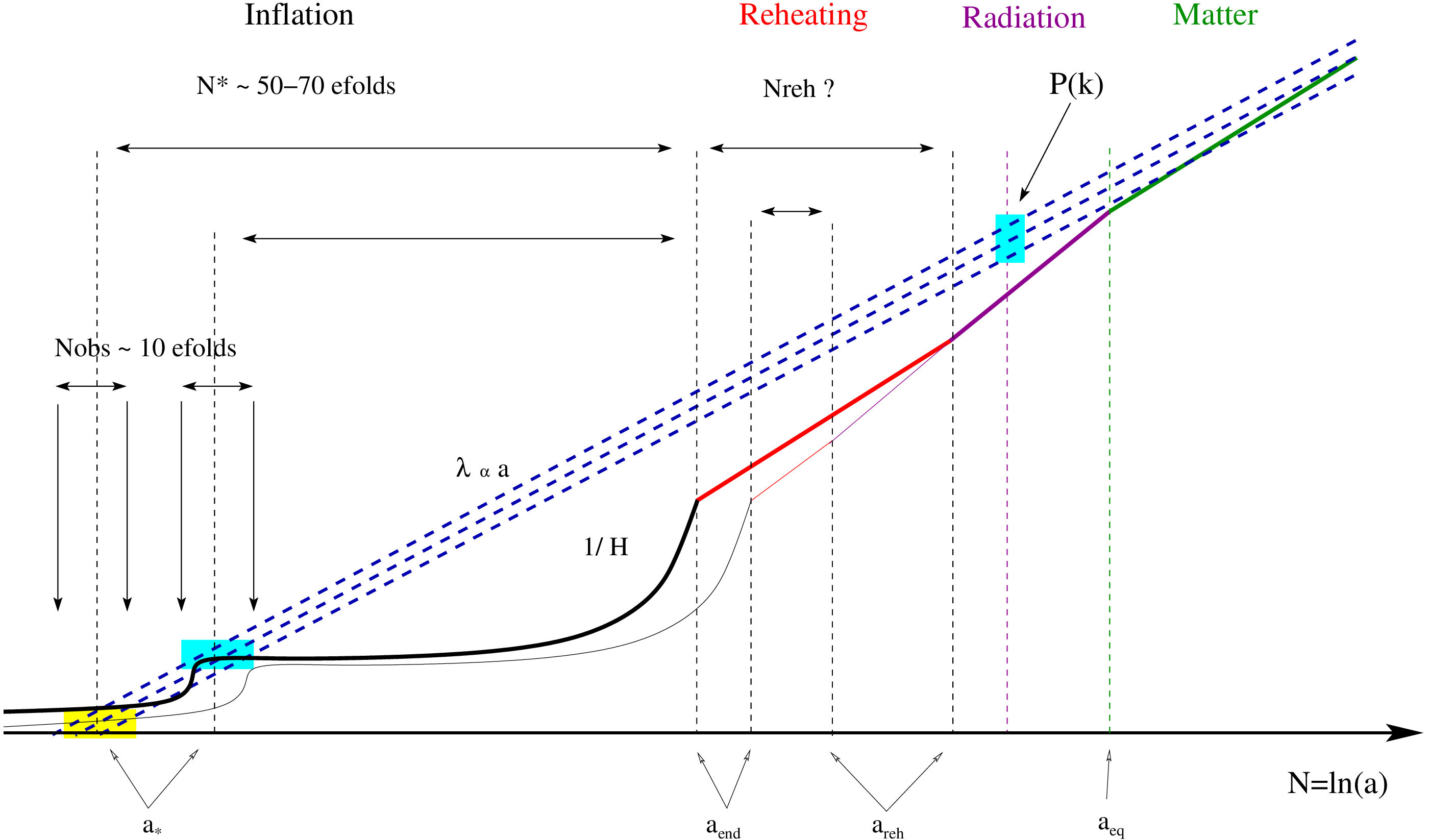}
\caption{Sketch of length scale evolution in inflationary
cosmology. The horizontal axis is the number of efold while the
vertical axis represents a logarithmic measure of lengths. The
cosmological stretching of the observable wavelengths is represented
by the three blue dashed lines. The evolution of the Hubble radius is
represented by a solid line from inflation to the matter era. In
between, the reheating era connects the end of inflation to the
radiation era. Although the redshift of equality is known, the
uncertainties existing on the reheating lead to uncertainties on the
redshift at which the observable wavelength today have left the Hubble
radius during inflation. As a result, even for a given model of
inflation, the resulting power spectra can be significantly different
if the number of efold during reheating is changed. This is
illustrated on this plot by the observability of an inflaton potential
feature that may accordingly be observable or not.}
\label{fig:scales}
\end{center}
\end{figure}

A great advantage of the slow-roll approach is that one does not need
to specify an explicit model of inflation~\cite{Martin:1999wa,
Liddle:1994dx, Martin:1997zd, Martin:2004um}. The constraints verified
by $\epsone$ and $\epstwo$ apply to all (single field) inflationary
models verifying the slow-rolling conditions $\epsone \ll 1$ and
$\epstwo \ll 1$ (see however
Refs.~\cite{Schwarz:2004tz,deOliveira:2005mf}). However, using these
results for a given model of inflation requires the knowledge of $N_*$
to determine the associated theoretical values of
$\epsilon_i(N_*)$. As can be seen in Fig.~\ref{fig:scales}, the value
of $N_*$ depends on the number of efold $N_\ureh$ during which the
universe reheated before the radiation era. The reheating era depends
on the microphysics associated with the decay of the inflaton field
whose complexity renders the determination of $N_\ureh$
difficult~\cite{Felder:2000hj, Kofman:1997yn, Garcia-Bellido:1997wm,
Senoguz:2004vu, Podolsky:2005bw, Desroche:2005yt,
Allahverdi:2006wh}. However, under reasonable assumptions, it has been
shown in Ref.~\cite{Liddle:2003as} that typically $40 \lesssim N_*
\lesssim 60$, although these bounds may vary by a factor of two for
extreme models. For the large field models represented in
Fig.~\ref{fig:srwmap}, the uncertainties on the reheating blur the
theoretically predicted values of the slow-roll parameters (see the
short segments in Fig.~\ref{fig:srwmap}). Notice that the resulting
errors in the $\epsilon_i$ remain small compared to the current CMB
data accuracy, but this is not necessarily the case for other models,
as for instance the small field models discussed in the following. The
problem is expected to become even more significant with the next
generation of more accurate CMB measurements.

Another difficulties that may show up with the slow-roll approach
concern the existence of features in the inflaton potential. Although
the current data support the almost scale invariance of the primordial
power spectra, the presence of sharp localised deviations is still
allowed by the data and might even be favoured~\cite{Martin:2006rs,
Barriga:2000nk, Martin:2003sg, Martin:2004iv, Martin:2004yi,
Easther:2004vq, Hunt:2004vt, Covi:2006ci}. In the framework of field
inflation, features in the power spectra generically result from
transient non-slow-rolling evolution associated with sharp features in
the inflation potential. In these cases, deriving analytical
approximations for the perturbations require the use of more involved
methods~\cite{Martin:2002vn, Casadio:2004ru, Casadio:2005xv,
Casadio:2006wb}. Let us also stress that one of the key ingredient
rendering analytical methods attractive is the conservation of the
comoving curvature perturbation $\curvpert$ on super-Hubble scales. As
illustrated in Fig.~\ref{fig:scales}, this allows one to identify the
scalar and tensor power spectra deep in the radiation era, which seed
the CMB anisotropies and structure formation, to the ones derived a
few efolds after Hubble exit during inflation. However, as soon as
inflation is driven by more than one field, the existence of
isocurvature modes that may source $\zeta$ after Hubble exit requires
that the modes evolution should be traced till the end of
reheating~\cite{Gordon:2000hv}. Although analytical methods can still
be used, their use is restricted by their domain of
validity~\cite{Noh:2001ia, DiMarco:2002eb,DiMarco:2005nq}.

The previous considerations suggest to use numerical methods to
directly compute the inflationary perturbations and deduce the
primordial power spectra. Numerical integrations in inflation are not
new and have been used to test the validity of analytical
approximations, or to derive the shape of the power spectra for some
particular models~\cite{Salopek:1988qh, Grivell:1999wc, Adams:2001vc,
Tsujikawa:2002qx, Parkinson:2004yx, Makarov:2005uh, Chen:2006xj}. However, with the
advent of MCMC methods in cosmology, one may be interested in merging
a full numerical integration of the perturbation during inflation with
the CMB codes such as $\camb$ and $\cosmomc$~\cite{Lewis:1999bs,
Lewis:2002ah}. The advantage of such an approach is that the
inflationary parameters become part of the cosmological model under
scrutiny. When compared with the data, one may expect to get
consistent marginalised constraints on both the parameters entering
the inflaton potential and the usual cosmological parameters
describing the radiation and matter content of the observed
universe. From a Bayesian point of view, this is the method that
should be used when one is interested in assessing the probability of
one model to explain the data (statistical evidence). Moreover, since
the underlying approximation is just linear perturbation theory, one
may consider without additional complications the treatment of
non-standard models as those involving several fields. Let us mention
that our objective is to use numerical methods in a well defined
theoretical framework. For a given model, we use the CMB data to
constrain the theoretical parameters and discuss the overall model
ability to explain the data. Numerical (non-exact) methods have also
been used in the context of the potential reconstruction problems
where the goal is to constrain the shape of the inflaton potential
along the observable window~\cite{Peiris:2006sj, Peiris:2006ug,
Kinney:2006qm} (see Fig.~\ref{fig:scales}).

However, as discussed before, to solve the cosmological perturbations
from their creation as quantum fluctuations during inflation to now,
it is necessary to model the reheating era. In fact, the importance of
reheating for inflation is very similar to the importance of the
reionisation for the CMB anisotropies. Although reionisation of the
universe is a complex process, its basic effects on the CMB
anisotropies can be modelled through the optical depth
$\optdepth$. Similarly, we will show that the basic effects induced by
the reheating on the inflationary perturbations may be taken into
account through a new parameter $\lnR$ which has not been considered
so far.

The plan is as follow. In a first section, the theoretical setup is
introduced. We use the sigma-model formalism which allows an easy
implementation of any scalar field inflation models in General Relativity
and multi-scalar tensor theories. The equations of motion for the
background fields and their perturbations are presented in the first
section. Their numerical integration is the subject of the second
part. After having introduced our modelisation of the reheating era,
the last section illustrates the usefulness of the exact numerical
method by an analysis of the third year WMAP data in the context of
the small field models

\section{Multifield inflation}

It is out of the scope of this work to deal with all the inflationary
models proposed so far. However, four-dimensional effective actions
associated with many inflation models, and especially the ones being
embedded in extra-dimensions, share the common feature that they
involve several scalar fields that may be non-minimally coupled to
gravity. This is for instance the case for the moduli associated with
the position of the branes in various string-motivated inflation
models~\cite{Brax:2004xh}. It is therefore convenient to consider an
action that may generically drive the dynamics of both minimally and
non-minimally coupled scalar fields, as in the
sigma-model~\cite{Damour:1992we, Damour:1993id, Koshelev:2005wk,
Ringeval:2005yn}.

\subsection{Sigma-model formalism}

Denoting by $\field^a(x^\mu)$ the $n_\sigma$ dimensionless scalar
fields living on a sigma-model manifold with metric
$\metric_{ab}(\field^c)$, we consider the action
\begin{equation}
\label{eq:actionsigmodel}
\begin{aligned}
S & = \dfrac{1}{2 \kappa^2} \int \bigg[R - \metric_{ab} g^{\mu \nu}
  \partial_\mu \field^a \partial_\nu \field^b -
   2\Vall(\field^c) \bigg] \sqrt{-g} \, \ud^4 x,
\end{aligned}
\end{equation}
where $g_{\mu\nu}$ is the usual four dimensional metric tensor of
determinant $g$, $R$ the Ricci scalar and $\Vall$ the field
potential.

For instance, if $\mat = \field^{(1)}/\kappa$ and $\metric_{11} = 1$,
this action describes a unique minimally coupled scalar field. In this
case, the associated potential is $\Vmat(\mat) = \Vall(\mat)/\kappa^2$
and we recover the standard form
\begin{equation}
\label{eq:actionsingle}
S = \dfrac{1}{2 \kappa^2} \int R \sqrt{-g} \, \ud^4 x + \int
 \left[-\dfrac{1}{2} g^{\mu \nu} \partial_\mu \mat \partial_\nu \mat -
 \Vmat(\mat) \right] \sqrt{-g} \, \ud^4 x.
\end{equation}
Another example is provided by the models of brane inflation where the
inflaton field $\mat$ lives on the brane and a bulk field $\moduli$ in
the four-dimensional effective action couples to gravity in a
non-minimal way~\cite{Langlois:2002bb, Maartens:2003tw,
Brax:2004xh}. With $n_\sigma=2$, $\mat = \field^{(1)}/\kappa$ and
$\moduli = \field^{(2)}$, Eq.~(\ref{eq:actionsigmodel}) can be recast
into
\begin{equation}
\label{eq:actionboundinf}
\begin{aligned}
  S &= \dfrac{1}{2 \kappa^2 } \int \left[R - g^{\mu \nu} \partial_\mu
      \moduli \partial_\nu \moduli - 2\Vgrav(\moduli) \right] \sqrt{-g}
      \,\ud^4 x \\ & + \int \left[-\dfrac{1}{2} \Afac^2(\moduli)
      \partial_\mu \mat \partial_\nu \mat -
      \Afac^4(\moduli)\Vmat(\mat) \right] \sqrt{-g} \, \ud^4 x ,
\end{aligned}
\end{equation}
for
\begin{equation}
\metric_{ab} = \diag(A^2,1), \qquad \Vall(\mat,\moduli) =
\Vgrav(\moduli) + \kappa^2 A^4(\moduli) \Vmat(\mat).
\end{equation}
This action describes the dynamics, in the Einstein frame, of the
field $\mat$ evolving in a potential $\Vmat$ in a scalar-tensor theory
of gravity where $\moduli$ is the scalar partner to the
graviton~\cite{Schimd:2004nq}. The conformal function $A^2(\moduli)$
and the self-interaction potential $W(\moduli)$ depend on the brane
setup considered~\cite{Lukas:1998tt, Lukas:1999yn, Brax:2000xk,
Kobayashi:2002pw}. In the general case, Eq.~(\ref{eq:actionsigmodel})
can be used to describe multifield inflation in a multi-scalar tensor
theory of gravity.

Differentiating the action (\ref{eq:actionsigmodel}) with respect to
the metric leads to the Einstein equations
\begin{equation}
\label{eq:einstein}
\begin{aligned}
G_{\mu \nu} =\source_{\mu\nu},
\end{aligned}
\end{equation}
with the source terms
\begin{equation}
\label{eq:tmunu}
\source_{\mu\nu}=\metric_{ab} \source^{ab}_{\mu \nu} -
g_{\mu\nu} \Vall,
\end{equation}
where
\begin{equation}
  \source^{ab}_{\mu \nu} = \partial_{\mu}\field^a \partial_{\nu}
  \field^b - \dfrac{1}{2}
  g_{\mu\nu} \partial_\rho \field^a \partial^\rho \field^b.
\end{equation}
Similarly, the fields obey the Klein-Gordon-like equation
\begin{equation}
\label{eq:kleingordon}
\square \field^c + g^{\mu \nu} \christoffel^c_{ab} \partial_\mu \field^a
\partial_\nu \field^b = \Vall^c,
\end{equation}
where $\christoffel$ denotes the Christoffel symbol on the field-manifold
\begin{equation}
\christoffel^c_{ab} = \dfrac{1}{2} \metric^{cd} \left(\metric_{da,b} +
  \metric_{db,a} - \metric_{ab,d} \right),
\end{equation}
and $\Vall^c$ should be understood as the vector-like partial
derivative of the potential
\begin{equation}
\Vall^{c} = \metric^{cd} \Vall_d = \metric^{cd} \dfrac{\partial
  \Vall}{\partial \field^d}.
\end{equation}

\subsection{Background evolution}

In a flat (FLRW) universe with metric
\begin{equation}
\label{eq:flrw}
\ud s^2 = g_{\mu \nu} \ud x^\mu \ud x^\nu = a^2(\eta) \left(-\ud \eta^2
+ \delta_{ij} \ud x^i \ud x^j \right),
\end{equation}
$\eta$ being the conformal time and $i$ and $j$ referring to the
spatial coordinates, the equations of motion (\ref{eq:einstein}) and
(\ref{eq:kleingordon}) simplify to
\begin{align}
\label{eq:einsteinTT}
3 \calH^2 & = \dfrac{1}{2} \metric_{ab} {\field}^{a}{}'
{\field}^{b}{}' +
a^2 \Vall , \\
\label{eq:einsteinIJ}
2 \calH' &+ \calH^2 = -\dfrac{1}{2} \metric_{ab} {\field^{a}}{}'
{\field^{b}}{}'+  a^2 \Vall, \\
\label{eq:modulicosmo}
{\field^c}{}'' &+ \christoffel^c_{ab} {\field^a}{}'
{\field^b}{}' +2 \calH {\field^c}{}' = - a^2 \Vall^c,
\end{align}
where a prime denotes differentiation with respect to the conformal
time and $\calH = a \hubble $ is the conformal Hubble parameter. In
terms of the efold time variable $n$, the field equations can be
decoupled from the metric evolution and one gets
\begin{align}
\label{eq:hubblesquare}
  &\hubble^2 = \dfrac{\Vall}{3 - \dfrac{1}{2}\dot{\adia}^2},\\
\label{eq:hubbledot}
  &\dfrac{\dot{\hubble}}{\hubble}  = -\dfrac{1}{2} \dot{\adia}^2, \\
\label{eq:fieldevol}
  &\dfrac{{\ddot{\field}^{c}} + \christoffel^c_{ab} \dot{\field}^{a}
    \dot{\field}^{b}}{3 - \dfrac{1}{2} \dot{\adia}^2}  +
  \dot{\field}^{c} = - \dfrac{\Vall^c}{\Vall},
\end{align}
a dot being a differentiation with respect to $n$. We have introduced
a velocity field
\begin{equation}
\label{eq:adiadot}
\dot{\adia} = \sqrt{\metric_{ab} \dot{\field}^{a} \dot{\field}^{b}}.
\end{equation}
In fact, $\adia$ is the so-called adiabatic field introduced in
Ref.~\cite{Gordon:2000hv} which describes the collective evolution of
all the fields along the classical trajectory. From
Eqs.~(\ref{eq:hubbledot}) and Eq.~(\ref{eq:fieldevol}) one may
determine its equation of motion
\begin{equation}
\label{eq:adiaevol}
\adia'' + 2 \calH \adia' + a^2
\Vall_\sigma = 0,
\end{equation}
with $\Vall_\sigma \equiv \unitv^c \Vall_c$ and where the $\unitv^a$
are unit vectors along the field trajectory:
\begin{equation}
\label{eq:unitvector}
\unitv^a \equiv \dfrac{{\field^a}{}'}{\adia'}
= \dfrac{\dot{\field}^{a}} {\dot{\adia}}.
\end{equation}
In the Einstein frame, inflation occurs for $\ud^2 a/\ud t^2 >0$, or
in terms of the first Hubble flow function, for\footnote{Notice that
this does not imply that the universe is accelerating in the string
frame and one has to verify that there are enough efolds of inflation
to solve the homogeneity and flatness issues in that
frame~\cite{Ringeval:2005yn, Esposito-Farese:2000ij, Martin:2005bp}.}
\begin{equation}
\label{eq:epsone}
\epsone \equiv -\dfrac{\dot{H}}{H} = \dfrac{1}{2} \dot{\sigma}^2  < 1.
\end{equation}
The multifield system induces an accelerated expansion of the universe
if the resulting adiabatic field velocity $\dot{\sigma}$ remains less
than $\sqrt{2}$. According to Eq.~(\ref{eq:fieldevol}), the term in
$1/(3 - \dot{\adia}^2/2)$ acts as a relativistic-like inertia for the
fields evolution and thus $\dot{\sigma} < \sqrt{6}$ (for a positive
potential). In this equation, the first term on the left hand side may
be interpreted as a covariant acceleration on the curved field
manifold, the second as a constant friction force and the right hand
side as a driving force deriving from the potential $\ln \Vall$. In
fact, we recover the well-known attractor behaviour of the
inflationary evolution: whatever the initial fields velocity, the
friction term ensures that the terminal velocity of the fields will
be, after a transient regime,
\begin{equation}
\label{eq:transient}
\dot{\field^a} \simeq -\dfrac{\ud \ln \Vall}{\ud \field^a}.
\end{equation}
Analytical integration of the previous expression is at the basis of
the slow-roll approximation when the effective potential $\ln \Vall$
is flat enough. In the general case, the driving force is always
pushing the fields towards the minimum of $\ln \Vall$. Let us note
that this is why the monomial potentials $\Vall \propto \mat^p$ are
actually ``flat'' for the large field values: $\ud \ln \Vall/\ud \mat
=p/\mat$ which goes to zero for $\mat \rightarrow \infty$.

\subsection{Linear perturbations}

\subsubsection{Scalar modes}
\label{sect:scalmodes}
In the longitudinal gauge, the scalar perturbations (with respect to
the rotations of the three-dimensional space) of the FLRW metric can
be expressed as
\begin{equation}
\label{eq:pertscalmetric}
\ud s^2 = a^2 \left[ - \left(1 + 2 \Phipert \right) \ud \eta^2 +
  \left(1 - 2 \Psipert \right) \gamma_{ij} \ud x^i \ud x^j \right],
\end{equation}
where $\Phipert$ and $\Psipert$ are the Bardeen potentials. With
$\fieldpert^a$ the field perturbations, the Einstein equations
perturbed at first order read
\begin{align}
\label{eq:energypert}
3 \calH \Psipert' & + \left(\calH'+2\calH^2 \right)\Psipert - \Delta
\Psi = - \dfrac{1}{2} \metric_{ab} {\field^a}{}' {\fieldpert^b}{}'
\nonumber \\ & - \dfrac{1}{2} \left(\dfrac{1}{2}\metric_{ab,c} {\field^a}{}'
  {\field^b}{}' + a^2 \Vall_c \right) \fieldpert^c,  \\
\label{eq:momentumpert}
\Psipert'& + \calH \Psipert = \dfrac{1}{2} \metric_{ab} {\field^a}{}'
\fieldpert^b, \\
\label{eq:diagonalpert}
\Psipert'' &+ 3\calH \Psipert' + \left(\calH' + 2\calH^2 \right)
\Psipert = \dfrac{1}{2} \metric_{ab} {\field^a}{}' {\fieldpert^b}{}'
\nonumber \\ & + \dfrac{1}{2} \left(\dfrac{1}{2} \metric_{ab,c} {\field^a}{}'
  {\field^b}{}' - a^2 \Vall_c \right) \fieldpert^c ,
\end{align}
where use has been made of $\Phipert=\Psipert$ from the perturbed
Einstein equation with $i\ne j$. For flat spacelike hypersurfaces,
\begin{equation}
\Delta \equiv \delta^{ij} \partial_i \partial_j.
\end{equation}
Similarly, the perturbed Klein-Gordon equations read
\begin{align}
  \label{eq:fieldpert} {\fieldpert^c}{}'' & + 2 \christoffel^c_{ab}
  {\field^a}{}' {\fieldpert^b}{}' + 2 \calH {\fieldpert^c}{}'
  \nonumber \\ & + \Bigg(\christoffel^c_{ab,d}
  {\field^a}{}'{\field^b}{}' + a^2\Vall^c_d -
  \metric^{ca}\metric_{ab,d} \, a^2 \Vall^b \Bigg) \fieldpert^d
  \nonumber \\ & - \Delta \fieldpert^c = 4 \Psipert' {\field^c}{}' -
  2\Psipert a^2 \Vall^c.
\end{align}
As discussed in the introduction, if there is more than one scalar
field involved, the entropy perturbation modes can source the
adiabatic mode even after Hubble exit. The equation governing the
evolution of the comoving curvature perturbation $\curvpert$ can be
obtained from Eqs.~(\ref{eq:energypert}) to (\ref{eq:diagonalpert}),
using the background equations. Firstly, the Bardeen potential
verifies
\begin{equation}
\label{eq:bardeenevol}
\begin{aligned}
\Psipert'' + 6 \calH \Psipert' + \left(2 \calH' + 4 \calH^2 \right)
\Psipert - \Delta \Psipert = - a^2 \Vall_c \fieldpert^c.
\end{aligned}
\end{equation}
Using the geometrical definition for the comoving curvature
perturbation~\cite{Martin:1997zd}
\begin{equation}
\label{eq:curvpertdef}
\curvpert \equiv \Psipert - \dfrac{\calH}{\calH' - \calH^2} \left(
\Psipert' + \calH \Phipert \right),
\end{equation}
Eq.~(\ref{eq:momentumpert}) yields
\begin{equation}
\label{eq:curvpertadiab}
\curvpert= \Psipert + \calH \dfrac{\adiapert}{\adia'},
\end{equation}
where the adiabatic perturbation $\adiapert$ is also the resulting
perturbation of all fields projected onto the classical trajectory
[see Eqs.~(\ref{eq:adiadot}) and (\ref{eq:unitvector})]:
\begin{equation}
\adiapert = \dfrac{\metric_{ab} {\field^a}{}'
  \fieldpert^b}{\adia'} = \unitv_a \fieldpert^a.
\end{equation}
The dynamical equation (\ref{eq:bardeenevol}) now exhibits couplings
between the adiabatic and entropy modes
\begin{equation}
\curvpert' = \dfrac{2\calH}{\adia'^2} \Delta \Psipert -
\dfrac{2\calH}{\adia'^2} \left(a^2 \Vall_a \fieldpert^a -
  a^2\dfrac{\Vall_c {\field^c}{}'}{\adia'} \dfrac{\metric_{ab}
    {\field^a}{}' \fieldpert^b}{\adia'} \right),
\end{equation}
which can be recast into
\begin{equation}
\label{eq:curvpertevol}
\curvpert' = \dfrac{2\calH}{\adia'^2} \Delta \Psipert -
\dfrac{2\calH}{\adia'^2}  \orthometric^c_d a^2 \Vall_c \fieldpert^d.
\end{equation}
The orthogonal projector is defined by
\begin{equation}
\orthometric_{ab} = \metric_{ab} - \firstform_{ab},
\end{equation}
where $\firstform_{ab}\equiv \unitv_a \unitv_b$ is the first
fundamental form of the one-dimensional manifold defined by the
classical trajectory~\cite{Carter:1997pb}. Clearly, the comoving
curvature perturbation on super-Hubble scales for which $\Delta
\Psipert \simeq 0$ is only sourced by the entropy perturbations
defined as the projections of all field perturbations on the
field-manifold subspace orthogonal to the classical trajectory. If, on
the contrary, there is a single field involved during inflation, then
these terms vanish and we recover that $\zeta$ remains constant after
Hubble exit.

\subsubsection{Tensor modes}

In the Einstein frame, the scalar and tensor degrees of freedom are
decoupled. Therefore, the equation of evolution for the tensor modes
remains the same as in General Relativity. For a flat perturbed FLRW
metric
\begin{equation}
\label{eq:perttensmetric}
\ud s^2 = - a^2 \ud \eta^2 + a^2 \left(\delta_{ij} +
\tensor_{ij}\right) \ud x^i \ud x^j,
\end{equation}
where $\tensor_{ij}$ is a traceless and divergenceless tensor
\begin{equation}
\delta^{ij} \tensor_{ij} = \delta^{ik} \partial_k \tensor_{ij} = 0,
\end{equation}
one gets~\cite{Mukhanov:1990me,Liddle:1993fq}
\begin{equation}
\label{eq:tensevol}
\tensor_{ij}'' + 2 \calH \tensor_{ij}' - \Delta \tensor_{ij}=0.
\end{equation}

\subsubsection{Primordial power spectra}

The initial conditions for the cosmological perturbations require the
knowledge of the two-point correlation functions for all of the
observable scalar and tensor modes deep in the radiation era. In
Fourier space, these are just the power spectra associated with the
values taken by the adiabatic and entropy perturbations at the end of
the reheating, \ie
\begin{equation}
\label{eq:scalpowerspect}
\begin{aligned}
  \power_{ab} & = \dfrac{\kwav^3}{2 \pi^2}
  \left[\obspert^{a}(\kwav)\right]^* \left[\obspert^{b}(\kwav)\right],
\end{aligned}
\end{equation}
where $\obspert^a$ stands for $\curvpert$ or the entropy
modes. Similarly, taking into account the polarisation degrees of
freedom, the tensor power spectrum reads
\begin{equation}
\label{eq:tenspowerspect}
\begin{aligned}
  \powerTens(\kwav) & = \dfrac{2 \kwav^3}{\pi^2} \left| \tensor(\kwav)
  \right|^2.
\end{aligned}
\end{equation}

In the following, we summarise the numerical method used to solve the
full set of Einstein and Klein-Gordon equations derived in this
section. The power spectra can then be deduced from
Eqs.~(\ref{eq:scalpowerspect}) and (\ref{eq:tenspowerspect}) by
pushing the integration till the end of the reheating.

\section{Numerical method}
\label{sect:numinf}
\subsection{Integrating the background}

As suggested by the form of Eqs.~(\ref{eq:hubblesquare}) to
(\ref{eq:fieldevol}), it is convenient to use the number of efold $n$
as the integration variable. In fact, the background evolution only
requires the integration of the fields equation of motion
(\ref{eq:fieldevol}). From the Cauchy theorem, the solution is unique
provided all the $\field^a(0)$ and $\dot{\field^a}(0)$ are
given at $n=0$. Plugging the solutions for $\field^a(n)$ into
Eq.~(\ref{eq:hubblesquare}) uniquely determine the Hubble parameter
and thus the geometry during inflation.

\subsubsection{Initial conditions}

However, as previously mentioned, the attractor behaviour induced by
the friction term erases any effect associated with the initial field
velocities after a few efolds. This is the very reason why initial
conditions in inflation are essentially related to the initial field
values only. On the numerical side, the attractor ensures the
stability of almost all forward numerical integration
schemes\footnote{as well as the instability of backward
integrations.}. In the following, we have used a Runge-Kutta
integration method of order five and the initial field velocities have
been chosen on the attractor by setting
\begin{equation}
\label{eq:attractor}
\dot{\field^a}(0) = - \left. \dfrac{\ud \ln \Vall}{\ud
\field^a} \right|_{\field^a(0)}.
\end{equation}

\begin{figure}
\begin{center}
\includegraphics[width=9cm]{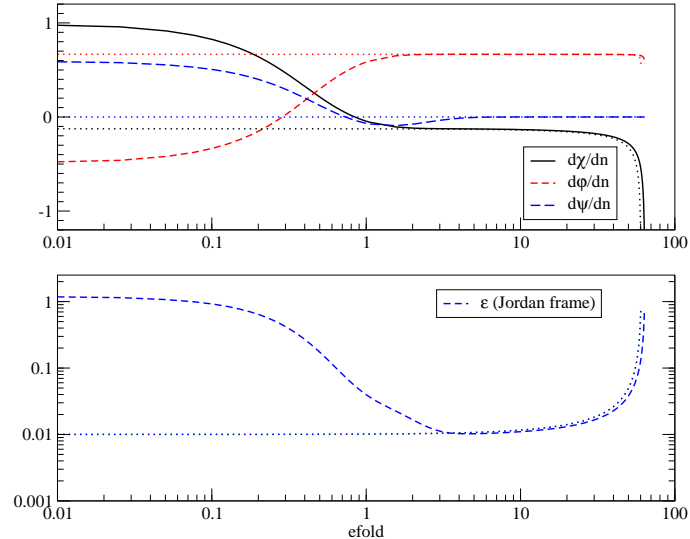}
\caption{Evolution of the field velocities $\dot{\field^a}
=\{\dot{\chi},\dot{\phi},\dot{\psi}\}$ in the boundary inflation model
of Ref.~\cite{Ringeval:2005yn} which involves three fields, two of
them being non-minimally coupled to gravity. The dotted curves are the
solutions obtained by setting the initial field velocities on the
attractor according to Eq.~(\ref{eq:attractor}) whereas the solid and
dashed curves are the solutions obtained from a random choice of the
initial field velocities (ensuring however $H^2>0$). The first Hubble
flow parameter in the string frame is plotted in the lower
panel. Since this is also the adiabatic field velocity squared, the
acceleration properties of the universe after a few efolds do not
longer depend on the initial field velocities [see
Eq.~(\ref{eq:epsone})].}
\label{fig:relax}
\end{center}
\end{figure}

The robustness of the attractor during inflation may be quantified by
comparing the numerical solutions obtained from various arbitrary
choices of the initial field velocities, at fixed value of
$\field^a(0)$. As an illustration, we have plotted in
Fig.~\ref{fig:relax} the efold evolution of $\dot{\field}^a(n)$ in a
brane inflation model involving three scalar fields, two of them are
non-minimally coupled to gravity and represent the position of two
branes in a five-dimensional bulk (see
Ref.~\cite{Ringeval:2005yn}). As can be seen on this plot, all the
fields are on the attractor after a few efolds.

\subsubsection{End of inflation}

\begin{figure}
\begin{center}
\includegraphics[width=9cm]{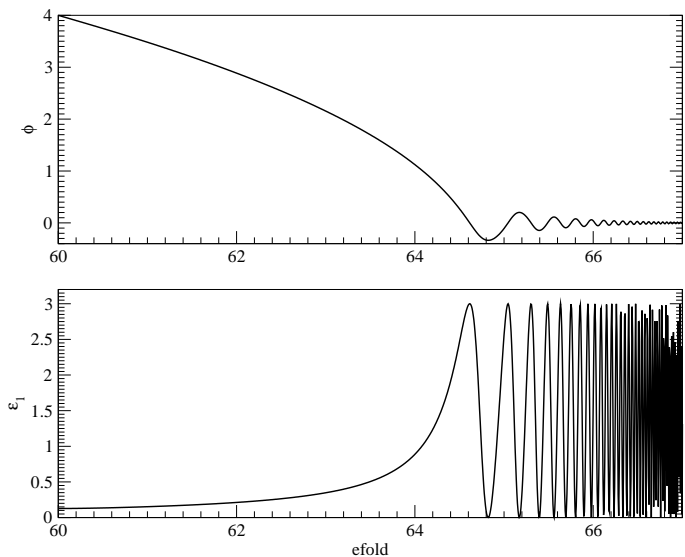}
\caption{End of inflation in the large field model $V \propto
  \mat^2$. The field oscillations around the minimum of the potential
  trigger its decay and the reheating era.}
\label{fig:endinf}
\end{center}
\end{figure}

From the above initial conditions the fields evolve toward the minimum
of the potential $\ln \Vall$ while the expansion of the universe
accelerates as long as $\epsone<1$. It would therefore be natural to
define the end of inflation by the efold $n_\uend$ at which
$\epsone(n_\uend)=1$. However, this is usually not the end of the
fields evolution since they have not yet reached the minimum of the
potential. On the contrary, $\epsone(n_\uend)=1$ just signals that the
kinetic terms in Eq.~(\ref{eq:actionsigmodel}) start to dominate over
the potential. Since the expansion factor is decelerating for
$\epsone(n_\uend)>1$, this late stage evolution takes place during a
few efolds and the fields rapidly reach the minimum of the
potential. In the standard picture, the fields oscillate around the
minimum of the potential and decay through parametric resonances into
the relativistic fluids present during the radiation
era~\cite{Kofman:1997yn, Turner:1983he} (see Fig.~\ref{fig:endinf})
The details of the reheating process are very model dependent and
require the knowledge of all the couplings between the inflaton and
the standard model particles~\cite{Bassett:2005xm}. This implies that
the number of efolds the universe reheated also depends on the model
at hand. Another complications come in multifield inflationary models
in which the end of inflation and the reheating may be triggered by
tachyonic instabilities. In these cases the condition
$\epsone(n_\uend)=1$ is not longer relevant and one should rather
introduce a limiting field value $\field_\uend$ to classically define
the beginning of the reheating era (and the end of inflation).

Following the previous discussion, our phenomenological approach to
the end of inflation is to assume an instantaneous transition to the
reheating era. The efold $n_\uend$ at which the transition occurs is
either determined by the condition $\epsone(n_\uend) = 1$ or when the
relevant field cross a limiting value $\field_\uend$; the choice being
made according to the inflation model we are interested in. The value
of $\field_\uend$ may be given by the underlying microphysics or
considered as an additional parameter of the inflation model. For
instance, as can be seen in Fig.~\ref{fig:endinf}, it is convenient to
define the end of inflation for the large field models by $\epsone=1$:
the field evolution afterwards, oscillations around the minimum of the
potential and subsequent decay, is supposed to be part of the
reheating stage.

Knowing the fields value $\field^a(n)$ from $n=0$ to $n=n_\uend$, the
background geometry is given by Eqs.~(\ref{eq:hubblesquare}) and
(\ref{eq:hubbledot}) and we can now numerically integrate the linear
perturbations on the same efold range. As discussed in the
introduction, the link with the cosmological perturbations observed
today still requires a reheating model that will be introduced in
Sect.~\ref{sect:reheating}.

\subsection{Integrating the perturbations}

As for the background, we have chosen to integrate the linear
perturbations in efold time. Focusing on the scalar perturbations,
their dynamics is driven by the Einstein and Klein-Gordon equations
given in Sect.~\ref{sect:scalmodes}. These ones are however redundant
due to the stress energy conservation already included in the Bianchi
identities. As a result, it is necessary to integrate only a subset of
Eqs.~(\ref{eq:energypert}) to (\ref{eq:fieldpert}). Although the
Bardeen potential $\Psipert$ could be explicitly expressed in terms of
the field perturbations $\fieldpert^a$ only, such an expression is
singular in the limit $\kwav \rightarrow 0$ and $\epsone \rightarrow
0$, which is not appropriate for a numerical integration (see
below). It is more convenient to simultaneously integrate the second
order equations (\ref{eq:fieldpert}) and
(\ref{eq:bardeenevol}). Recast in efold time, they read
\begin{align}
 \label{eq:fieldpertefold}
\ddot{\fieldpert^c} & + (3 - \epsone) \dot{\fieldpert^c} + 2
\christoffel^c_{ab} \dot{\field^a} \dot{\fieldpert^b} +
\Bigg(\christoffel^c_{ab,d} \dot{\field^a} \dot{\field^b} +
\dfrac{\Vall^c_d}{\hubble^2} - \metric^{ca}\metric_{ab,d}
\dfrac{\Vall^b}{\hubble^2} \Bigg) \fieldpert^d \nonumber \\ & +
\dfrac{\kwav^2 }{a^2 \hubble^2}\fieldpert^c = 4 \dot{\Psipert}
\dot{\field^c} - 2 \Psipert \dfrac{\Vall^c}{\hubble^2},\\
\label{eq:bardeenevolefold}
\ddot{\Psipert} & + (7 - \epsone) \dot{\Psipert} + \left(2
 \dfrac{\Vall}{\hubble^2} + \dfrac{\kwav^2}{a^2 \hubble^2} \right)
 \Psipert = - \dfrac{\Vall_c}{\hubble^2} \fieldpert^c.
\end{align}
The constraint equations (\ref{eq:energypert}) and
(\ref{eq:momentumpert}) being first integrals of the above equations,
there is still an integration constant that should be set to restore
the equivalence to the full set of Einstein and Klein-Gordon
equations. This one can be fixed by choosing the appropriate initial
conditions at $n=n_\uic$ for the Bardeen potential $\Psipert$. Setting
all the $\fieldpert^a(n_\uic)$ and $\dot{\fieldpert^a}(n_\uic)$, the
initial conditions for the Bardeen potential are indeed uniquely given
by Eq.~(\ref{eq:energypert}) and (\ref{eq:momentumpert}). In efold
time, one gets
\begin{equation}
\label{eq:bardeenic}
\begin{aligned}
\Psipert & = \dfrac{1}{2 \left(\epsone - \dfrac{\kwav^2}{a^2
    \hubble^2}\right)} \left[\metric_{ab} \dot{\field^a}
    \dot{\fieldpert^b} + \left( \dfrac{1}{2} \metric_{ab,c}
    \dot{\field^a}\dot{\field^b} + 3 \metric_{ac} \dot{\field^a} +
    \dfrac{\Vall_c}{\hubble^2} \right) \fieldpert^c\right],\\
    \dot{\Psipert} & = \dfrac{1}{2} \metric_{ab} \dot{\field^a}
    \fieldpert^b - \Psipert,
\end{aligned}
\end{equation}
these expressions being evaluated at the initial efold time. As a
result, the linear perturbations of both the fields and metric are
uniquely determined by the initial conditions $\fieldpert^a(n_\uic)$
and $\dot{\fieldpert^a}(n_\uic)$. As discussed in the next section,
the initial conditions are set on sub-Hubble scales for which $\kwav
\gg a \hubble$ ensuring the regularity of Eqs.~(\ref{eq:bardeenic}).

\subsubsection{Quantum initial conditions}

In the context of single field inflation, the initial conditions for
the linear perturbations are given by the quantum fluctuations of the
field-metric system on sub-Hubble scales $\kwav \rightarrow
\infty$. In this limit, the perturbations decouple from the expansion
of the universe and a field quantisation can be performed along the
lines described in Refs.~\cite{Mukhanov:1990me, Martin:2004um}. The
canonically normalised quantum degrees of freedom are encoded in the
Mukhanov-Sasaki variable
\begin{equation}
\label{eq:mukhadef}
\mukha = \adiapert +  \dfrac{\adia'}{\calH} \Psipert = \adiapert +
\sqrt{2 \epsone} \Psipert,
\end{equation}
with $\adia = \field^{(1)} = \kappa \varphi$ for single field models.
In terms of $\mukha$, the equation of motion (\ref{eq:fieldpert}) can
be recast into
\begin{equation}
(a \mukha)'' + \left[\kwav^2 - \dfrac{\left(a \sqrt{\epsone}
\right)''}{a \sqrt{\epsone}} \right] a \mukha = 0,
\end{equation}
showing that in the small scales limit $\kwav \rightarrow \infty$, the
quantity $a \mukha$ follows the dynamics of a free scalar
field. Assuming a Bunch-Davies vacuum, $a \mukha$ has a positive
energy plane wave behaviour on small scales and in Fourier space one
gets
\begin{equation}
\label{eq:bunchdavies}
\lim _{\kwav \rightarrow +\infty } a\mukha (\eta)=\kappa
\dfrac{\ue^{-i\kwav \eta }}{\sqrt{2 \kwav}}.
\end{equation}
This solution uniquely determines the subsequent evolution of the
perturbations during inflation and will be our starting point for the
numerical integration. According to Eq.~(\ref{eq:bardeenic}), provided
the initial conditions are set in the limit $\kwav/\calH \rightarrow
\infty$, Eq.~(\ref{eq:bunchdavies}) is also the small scales behaviour
of the rescaled adiabatic field perturbations $a \adiapert$.

For a multifield system the previous results can be generalised and
the quantum modes identify with the adiabatic perturbations
$\adiapert$ together with the canonically normalised entropy modes
introduced in Sect.~\ref{sect:scalmodes}. In fact, if the original
fields $\field^a$ are already canonically normalised, \ie
$\metric_{ab} = \delta_{ab}$, and independent dynamical variables in
the small scales limit, the adiabatic and entropy perturbations can be
obtained from the original field perturbations by local rotations on
the $n_\sigma$-dimensional field manifold~\cite{Gordon:2000hv,
DiMarco:2002eb, Ringeval:2005yn}. In this case, denoting by
$\entropert^a$ the adiabatic and entropy perturbations, with the
convention $\entropert^{(1)} = \adiapert$, one has
\begin{equation}
\entropert^a = \calM^a_b(\field^c) \fieldpert^b,
\end{equation}
where $\calM$ is an instantaneous rotation matrix, $\calM^\dag =
\calM^{\uT} = \calM^{-1}$, depending on the background quantities
only. Under these assumptions, the quantum modes are independent in
the small scales limit and their two point correlators reduce to
\begin{equation}
\left \langle {\entropert^a}^*(\vec{\kwav}) \entropert^b(\vec{\kwav'})
  \right \rangle \underset{\kwav \gg \calH}{=} \delta^{ab}
  \calP_\varsigma(\kwav) \delta(\vec{\kwav}-\vec{\kwav'}),
\end{equation}
$\calP_\varsigma$ being the free field power spectrum given by the square
modulus of Eq.~(\ref{eq:bunchdavies}). Consequently, all the
correlators between the original field perturbations inherit these
initial conditions:
\begin{equation}
\left \langle {\fieldpert^a}^*(\vec{\kwav}) \fieldpert^b(\vec{\kwav'})
\right \rangle = {\left(\calM^{-1}\right)^a_c}^*
\left(\calM^{-1}\right)^b_d \left \langle {\entropert^c}^*
\entropert^d \right \rangle \underset{\kwav \gg \calH}{=} \delta^{ab}
\calP_\varsigma(\kwav) \, \delta(\vec{\kwav}-\vec{\kwav'}).
\end{equation}
The previous results can be generalised for the sigma-models with a
diagonal metric $\metric_{ab}$ non equal to the identity by the
transformation $\fieldpert^a \rightarrow \sqrt{\metric_{aa}}
\fieldpert^a$ (no summation). In the general case, the transformation
matrix $\calM$ would mix all the fields and the cross
correlators. However, since it is always possible to diagonalise
$\metric_{ab}$ through a field redefinition, we will now assume
without lost of generality that $\metric_{ab}$ is diagonal.

Defining the normalised quantum modes by
\begin{equation}
\qmodeS^a = a \sqrt{2} \metric^{1/2}_{aa} \kwav^{3/2} \fieldpert^a,
\end{equation}
with $\metric_{aa}$ evaluated along the background solution, in the
small scales limit Eq.~(\ref{eq:bunchdavies}) can be recast into the
initial conditions
\begin{equation}
\label{eq:qmodeini}
\begin{aligned}
\qmodeS^a \underset{\kwav \gg \calH}{=} \kappa \kwav, \quad
\dfrac{{\qmodeS^a}'}{\kwav} \underset{\kwav \gg \calH}{=} -i \kappa \kwav,
\end{aligned}
\end{equation}
up to a phase factor. In terms of the field perturbations, the initial
conditions in efold time therefore read
\begin{equation}
\label{eq:fieldpertic}
\begin{aligned}
\left. \sqrt{2} \, \kwav^{3/2} \fieldpert^a\right|_\uic &= \dfrac{\kappa
\kwav}{a_\zero} \dfrac{a_\zero}{a_\uend}
\dfrac{\ue^{n_\uend-n_\uic}}{\sqrt{\metric_{aa}(n_\uic)}}\, ,\\
\left. \sqrt{2}\, \kwav^{3/2} \dot{\fieldpert^a} \right|_\uic &= - \dfrac{\kappa
\kwav}{a_\zero} \dfrac{a_\zero}{a_\uend}
\dfrac{\ue^{n_\uend-n_\uic}}{\sqrt{\metric_{aa}(n_\uic)}} \left[ 1 +
\dfrac{1}{2}
\dfrac{\dot{\metric_{aa}}(n_\uic)}{\metric_{aa}(n_\uic)} + i
\dfrac{\kwav}{a_\uic \hubble_\uic} \right].
\end{aligned}
\end{equation}
The efold $n_\uic$ at which these initial conditions should be set has
not been specified yet. In fact, the limit $\kwav/\calH \rightarrow
\infty$ would correspond to the infinite past and does not make sense
for non-eternal field inflation models\footnote{Eternal inflation may
occur when the quantum fluctuations on Hubble length scales become
dominant over the classical field evolution and the semi-classical
approach used here would no longer be valid, at least in the
self-reproducing regime.}. However, by definition of
$n=\ln{(a/a_\uini)}$, the condition $\kwav/a \gg \hubble$ is already
satisfied a few efolds before Hubble exit. For all inflation models
lasting more than $N_*$ efolds, it would be natural to set the initial
conditions for the perturbations at the beginning of inflation $n_\uic
= 0$ (see Fig.~\ref{fig:scales}).

Choosing $n_\uic=0$ is however not appropriate for a numerical
integration. Indeed, according to the initial values of the background
fields, the total number of efolds $n_\uend$ can be much greater than
$N_*$. In such cases, most of the computing time for the perturbations
would be spent into the deep sub-Hubble regime for which the modes
behave as free plane waves. It is rather more convenient to set the
initial conditions ``closer'' to the time $n_\kwav$ at which a given
mode cross the Hubble radius $\kwav = \calH(n_\kwav)$. Following
Ref.~\cite{Salopek:1988qh}, a simple choice is to define $n_\uic$ for
each mode according to
\begin{equation}
\label{eq:initime}
\dfrac{\kwav}{\calH(n_\uic)} = \constdec,
\end{equation}
$\constdec$ being a constant verifying $\constdec \gg 1$ and
characterising the decoupling limit. Strictly speaking, this choice
introduces small trans-Planckian-like interferences between the
modes\footnote{In the free field limit, a more accurate choice for
$n_\uic$ is $\kwav \eta(n_\uic) = \constdec$. This definition
maintains the phase factor in Eq.~(\ref{eq:bunchdavies}) independent
of $\kwav$ on the initial hypersurface.} which remain however
negligible provided $\constdec$ is big enough~\cite{Niemeyer:2002kh}.

\subsubsection{Mode integration}

For each perturbation mode of wavenumber $\kwav$,
Eqs.~(\ref{eq:fieldpertefold}) and (\ref{eq:bardeenevolefold}) are
numerically solved by setting the initial conditions
(\ref{eq:fieldpertic}) at the efold $n_\uic$, solution of
Eq.~(\ref{eq:initime}). In order to significantly speed-up the
numerical integration, instead of using the already computed
background solution it is more convenient to integrate both the
background and the perturbations simultaneously\footnote{For direct
numerical integrations, each step requires various forward and
backward evaluations of the background functions. If these ones are
not analytically known but precomputed, one has to use spline and
interpolation methods which are heavily time-consuming.}. This can be
done along the following steps.
\begin{figure}
\begin{center}
\includegraphics[width=9cm]{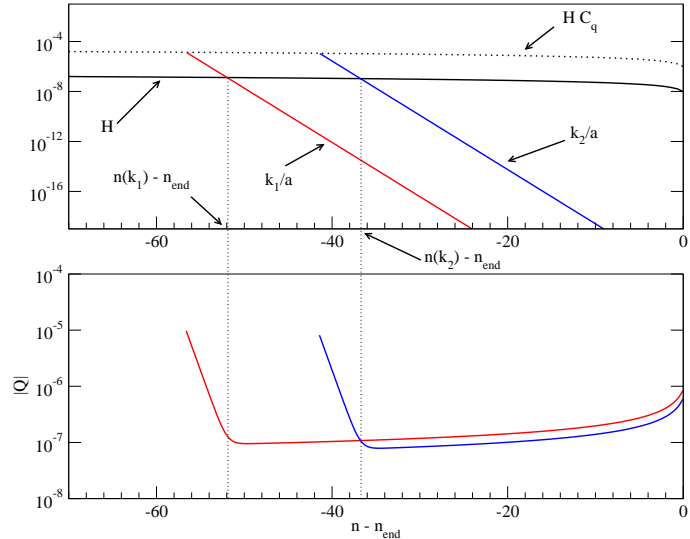}
\caption{Sketch of the numerical integration for the scalar
  perturbation in the $m^2 \mat^2$ single field model. In the top
  frame, the initial conditions for each mode are set at the efold
  $n(\kwav_i)$ solution of Eq.~(\ref{eq:initime}). The perturbations
  (and the background) are then integrated till the end of inflation
  at $n=n_\uend$. The bottom frame represents the efold evolution of
  the Mukhavov-Sasaki variable $\mukha$ for the two corresponding
  modes [see Eq.~(\ref{eq:mukhadef})]. Notice that only $\curvpert =
  \mukha/\sqrt{2\epsone}$ is conserved after Hubble exit.}
\label{fig:integsketch}
\end{center}
\end{figure}

Firstly, Eq.~(\ref{eq:initime}) is solved to determine
$n_\uic(\kwav_1)$ for the largest wavelength mode $\kwav_1$ we are
interested in. The background equations (\ref{eq:fieldevol}) and are
then integrated from $n=0$ to $n=n_\uic(\kwav_1)$. At that efold,
Eqs.~(\ref{eq:fieldevol}), (\ref{eq:fieldpertefold}) and
(\ref{eq:bardeenevolefold}) are simultaneously integrated till the end
of inflation at $n=n_\uend$. This process is iterated for each of the
$\kwav_i>\kwav_{i-1}$ mode wanted. However, to speed-up the
integration, it is enough to re-integrate the background from
$n_\uic(\kwav_{i-1})$ to $n_\uic(\kwav_{i})$ rather than from $n=0$
before switching on the perturbations (see
Fig.~\ref{fig:integsketch}). Such an integration gives the value of
all the field and metric perturbations at the end of inflation
$n=n_\uend$. In principle, the power spectra can then be deduced by
using Eq.~(\ref{eq:scalpowerspect}).

\subsubsection{Primordial power spectra}

For a multifield system, since the perturbation modes are supposed to
be independent deep under the Hubble radius, they can be considered,
from a classical point of view, as independent stochastic
variables. As a result, the power spectra at the end of inflation are
no longer given by Eq.~(\ref{eq:scalpowerspect}) but should be
computed as~\cite{Tsujikawa:2002qx}
\begin{equation}
\label{eq:numscalpowerspect}
\begin{aligned}
  \power_{ab} & = \dfrac{\kwav^3}{2 \pi^2} \sum_{m=1}^{n_\sigma}
  \left[\obspert_m^{a}(\kwav)\right]^* \left[\obspert_m^{b}(\kwav)\right],
\end{aligned}
\end{equation}
where, as before, $\obspert^a$ stands for the observable perturbations
one is interested in. The $\obspert^a$ can be the field perturbations
themselves but it is more customary for CMB analysis to use the
comoving curvature perturbation $\curvpert$ and the rescaled entropic
perturbations $\entropert^{a}/\dot{\adia}$ ($a>1$). The index ``$m$''
in Eq.~(\ref{eq:numscalpowerspect}) refers to the $n_\sigma$
independent initial conditions obtained by setting only one
perturbation mode $\qmodeS^m$ in the Bunch-Davies vacuum at $n_\uic$,
the other $\qmodeS^{q\ne m}$ vanishing. Notice that we have not
explicitly written the entropy modes since various definitions are
used in the literature. The definition of the entropy modes through
the standard orthogonalisation procedure along the field trajectory
can be found in Refs.~\cite{Gordon:2000hv,DiMarco:2002eb} and has the
advantage to give canonically normalised perturbations. Another
definitions introduce a reference field and define the entropy
perturbations to be the relative perturbations of the other fields
with respect to it. These differences come from the fact that one has
to specify how the fields decay after inflation to know between which
cosmological fluids entropy perturbations may exist. For instance, if
all the cosmological fluids observed today are produced by the decay
of one field only, then, although entropy perturbations exist during
inflation, they are usually not observable
afterwards~\cite{Weinberg:2004kf, Lemoine:2006sc}. Their only effect
would be to break the conservation of $\curvpert$ on super-Hubble
scale thereby requiring the integration of all the perturbations till
the end of inflation to determine the $\curvpert$ power spectrum.
\begin{figure}
\begin{center}
\includegraphics[width=9cm]{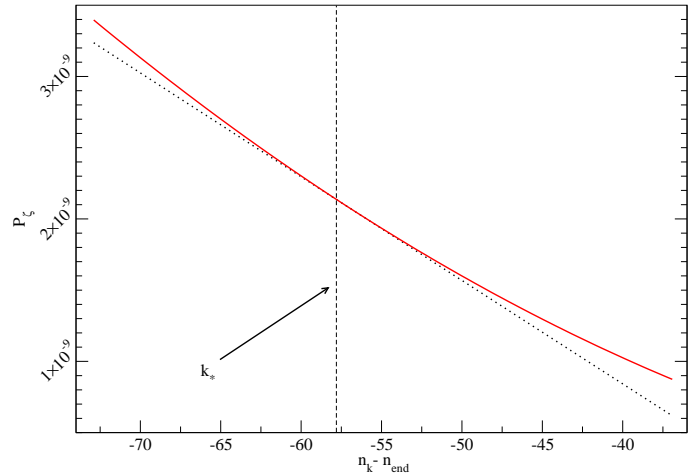}
\caption{Power spectra of the comoving curvature perturbation $\zeta$
  at the end of inflation from the first order slow-roll approximation
  (dotted line) and an exact numerical integration ($V \propto
  \mat^2$). The wavenumbers are expressed through $n_\kwav$, the efold
  at which the mode $\kwav$ crosses the Hubble radius:
  $\kwav=\calH(n_\kwav)$ (see Fig.~\ref{fig:integsketch}).}
\label{fig:compsrnum}
\end{center}
\end{figure}

\begin{figure}
\begin{center}
\includegraphics[width=9cm]{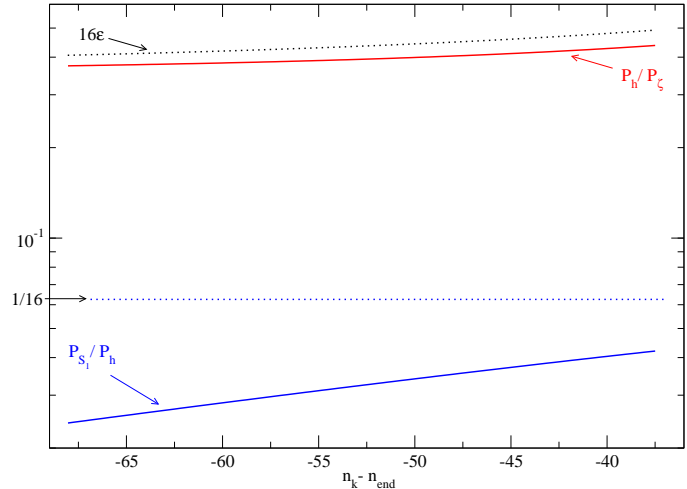}
\caption{Violation of the ``consistency check'' of inflation in a
non-minimally coupled two-fields model. The action is given by
Eq.~(\ref{eq:actionboundinf}) with $\Vgrav=0$, $\Vmat \propto \mat^2$
and $\Afac^2=\exp(-\alpha \moduli)$. Even for a small value of $\alpha =
1/30$, the ratio between the scalar and tensor power spectra is no
longer equal to $16 \epsone$. If the two fields were uncoupled on all
scales, then the power spectrum of entropy perturbations would be the
one of a free test scalar field $\calP_{S_1}=\calP_\tensor/16$. As can
be seen on the plot, this last condition is also violated.}
\label{fig:constcheck}
\end{center}
\end{figure}

As an illustration, Fig.~\ref{fig:compsrnum} shows the exact numerical
power spectrum for the comoving curvature perturbation $\zeta$ for the
single field chaotic model $\Vall \propto \mat^2$. As can be seen on
this plot, although the exact power spectrum differs from its first
slow-roll approximated version given by Eq.~(\ref{eq:srpowerscal}),
the differences remain small on a $10$ efold observable range.
Another example involving entropy perturbations is plotted in
Fig.~\ref{fig:constcheck} for a two fields model of inflation. The
presence of one entropy mode breaks the conservation of $\zeta$ on
super-Hubble scale and the so-called consistency check of inflation
$\calP_\tensor = 16 \epsone \calP_\zeta$ [see
Eqs.~(\ref{eq:srpowerscal}) and (\ref{eq:srpowertens}) at zero order].

\subsubsection{Tensor perturbations}

Since in the Einstein frame the tensor and scalar degrees of freedom
are decoupled, the numerical integration of the tensor modes does not
present any difficulties. The equation of motion (\ref{eq:tensevol})
can be recast into the equation of a parametric oscillator by defining
the canonical mode function $\qmodeT=k^{3/2} a \tensor$ satisfying the
deep sub-Hubble initial conditions (\ref{eq:bunchdavies}) for a
Bunch-Davies vacuum~\cite{Mukhanov:1990me}. The power spectrum
$\calP_\tensor$ is readily obtained by evaluating
Eq.~(\ref{eq:tenspowerspect}) at the end of inflation.

\subsubsection{Physical wavenumbers}

Up to now, one may have noticed that all the power spectra have been
plotted with respect to $n_\kwav$ and not with respect to the values
of $\kwav$. For astrophysical purpose, one needs to know the
correspondence between the comoving $\kwav$ appearing in the above
equations and the physical wavenumbers measured today $\kwav/a_0$,
whose typical unit is the $\Mpc^{-1}$. As it appears in the initial
conditions (\ref{eq:fieldpertic}), rendering $\kwav/a_0$ explicit
requires the knowledge of $a_0/a_\uend$, \ie the redshift $z_\uend$
associated with the end of inflation. As discussed in
Sect.~\ref{sect:intro}, this can only being done if one knows the
number of efolds during which the universe reheated. Let us also
notice that the physics involved in the quantum generation of
cosmological perturbations appears through these very numbers: $\kappa
\kwav/a_0$ are the wavenumbers measured today, usually of $\Mpc^{-1}$
size, expressed in unit of the Planck mass with $\kappa \equiv \sqrt{8
\pi}/\mpl$ [see Eq.~(\ref{eq:fieldpertic})].

\section{Application to CMB data analysis}

Fig.~\ref{fig:scales} makes clear that from the integration of the
perturbations described in the previous section, their power spectra
are known at the end of inflation. These primordial correlations then
evolve through the reheating, radiation and matter era to shape the
universe into its current state. The theory of cosmological
perturbations precisely predicts how such linear perturbations evolve
in a FLRW universe from the tightly coupled regime deep inside the
radiation era to today. As a result, we still have to know how the
power spectra are modified through the reheating era. As already
mentioned, reheating is very model dependant and a detailed analysis
is out of the scope of our current approach. Instead, remembering that
the objective is to use the CMB anisotropies measurements as a probe
to get information on the primordial correlations, we introduce a
basic model of reheating described by some phenomenological
parameters.

\subsection{Reheating}
\label{sect:reheating}

Assuming that perturbations on super-Hubble scales are not
significantly modified till the beginning of the radiation
era\footnote{Although such an assumption is motivated by the fact that
the physical processes involved during reheating are sub-Hubble, this
assumption may not longer be true in presence of entropy modes.}, the
reheating may influence the observed power spectra through its effects
on $z_\uend$ (see Fig.~\ref{fig:scales}). For instantaneous
transitions between inflation, the reheating era and the radiation
era~\cite{Liddle:2003as}, one has
\begin{equation}
\label{eq:zend}
\ln \dfrac{a_\uend}{a_\zero} = \ln \dfrac{a_\uend}{a_\ureh} + \ln
\dfrac{a_\ureh}{a_\ueq} + \ln \dfrac{a_\ueq}{a_\zero},
\end{equation}
where $a_\uend$, $a_\ureh$ and $a_\ueq$ are respectively the scale
factor at the end of inflation, at the end of reheating and at
equality between the energy density of radiation and the energy
density of matter. The redshift of equality can be expressed in terms
of the density parameter of radiation today $\OmegaR$ and the Hubble
parameter today $\hubble_\zero$. Moreover, during the radiation era
$\rho \propto a^{-4}$, and Eq.~(\ref{eq:zend}) can be recast into
\begin{equation}
\ln \dfrac{a_\uend}{a_\zero} = \ln \dfrac{a_\uend}{a_\ureh} -
\dfrac{1}{4} \ln \left(\kappa^4 \rho_\ureh\right) + \dfrac{1}{2} \ln
\left(\sqrt{3 \OmegaR} \kappa \hubble_\zero \right),
\end{equation}
where $\rho_\ureh$ denotes the total energy density at the end of the
reheating era. It is clear that the first two terms depend on the
physics involved during the reheating. For instance, they would only
depend on the energy density at the end of inflation for a
radiation-like reheating era. This suggests to introduce a
phenomenological parameter~\cite{Martin:2006rs}
\begin{equation}
\label{eq:defRrad}
\ln R_\urad \equiv \ln \dfrac{a_\uend}{a_\ureh} - \dfrac{1}{4} \ln
\left(\dfrac{\rho_\ureh}{\rho_\uend}\right).
\end{equation}
From this parameter, the quantity $\kwav/\calH$ entering the equations
of motion for the perturbations (\ref{eq:fieldpertefold}) and
(\ref{eq:bardeenevolefold}) can be evaluated in terms of the
$\kwav/a_0$ values measured today
\begin{equation}
\dfrac{\kwav}{a \hubble} = \dfrac{\kappa \kwav}{a_\zero}
\dfrac{1}{\kappa \hubble(n)} \dfrac{ \kappa \rho_\uend^{1/4}
  \,\ue^{n_\uend - n} }{R_\urad \left(3 \OmegaR\right)^{1/4}
  \sqrt{\kappa \hubble_\zero}}\,,
\end{equation}
and similarly for the initial conditions in
Eq.~(\ref{eq:fieldpertic}).

\subsection{CMB anisotropies}

For a given value of $R_\urad$, the primordial power spectra deep in
the radiation era are now uniquely determined by the numerical
integration described in Sect.~\ref{sect:numinf} and can be used as
initial conditions for the subsequent evolution of the perturbations.
\begin{figure}
\begin{center}
\includegraphics[width=9cm]{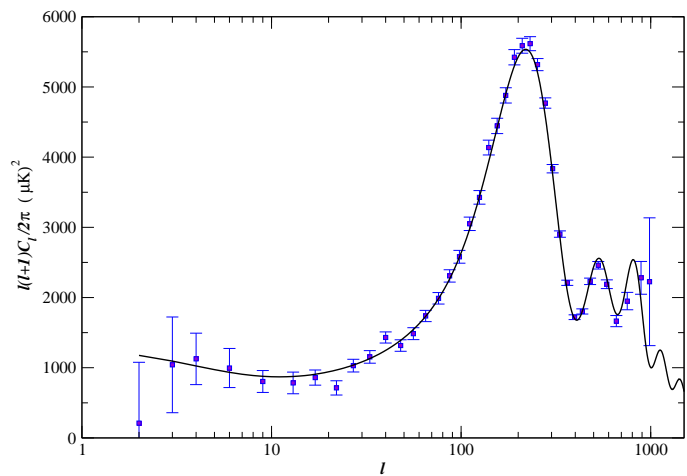}
\caption{Angular temperature power spectrum of the CMB anisotropies in
  a $\Lambda$CDM universe born under chaotic inflation. Fiducial
  values of the parameters have been used: $R_\urad =1$, $p=2$,
  $\kappa M = 2\times 10^{-3}$, $\OmegaB h^2 = 0.022$,
  $\OmegaCDM=0.12$, $h=0.7$, $z_\ure=12$, where $h$ is the reduced
  Hubble parameter and $z_\ure$ the redshift of reionisation. The WMAP
  third year measurements are represented as blue squares.}
\label{fig:clchaotic}
\end{center}
\end{figure}
The integration of the cosmological perturbations through the
radiation and matter era, as well as the resulting CMB anisotropies,
have been performed by using a modified version of the $\camb$
code~\cite{Lewis:1999bs}. The model parameters involved are both the
inflation parameters and the usual cosmological parameters describing
the FLRW model at late time. For instance, for a $\Lambda$CDM universe
experiencing large field inflation in its earliest times, there are
two parameters fixing the potential $\Vall(\mat) = M^4 \mat^p$, one
parameter describing the reheating era $R_\urad$, plus the four
cosmological base parameters: the number density of baryons $\OmegaB$,
of cold dark matter $\OmegaCDM$, the Hubble parameter today
$\hubble_\zero$ and the redshift of reionisation of the universe
$z_\ure$~\cite{Lewis:2002ah}. Let us recap that we have defined the
end of large field inflation by $\epsone(\mat_\uend)=1$. Combined with
the existence of the attractor during inflation, this ensures that
$\mat_\uend$ is fixed by the potential parameters [see
Eq.~(\ref{eq:transient})]. The resulting angular power spectrum for
the CMB temperature fluctuations is represented in
Fig.~\ref{fig:clchaotic} for a fiducial set of the parameters.

The next step is to use CMB measurements to constrain the models. For
this purpose, the parameter space can be sampled by using Markov Chain
Monte Carlo (MCMC) methods as implemented in the $\cosmomc$
code~\cite{Lewis:2002ah} to extract the probability distributions
satisfied by the model parameters. In the next section, we illustrate
such a procedure for the $\Lambda$CDM model born under small field
inflation by using the WMAP third year data~\cite{Page:2006hz,
Hinshaw:2006ia, Jarosik:2006ib}.

\subsection{WMAP3 constraints on small field models}

\begin{figure}
\begin{center}
\includegraphics[width=5.8cm]{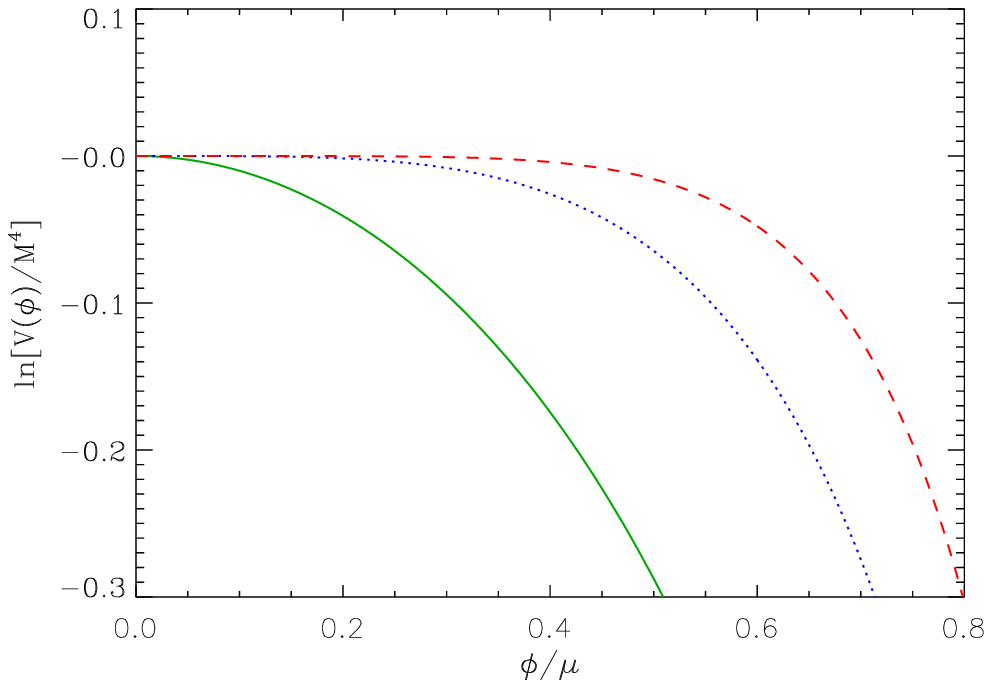}
\includegraphics[width=5.8cm]{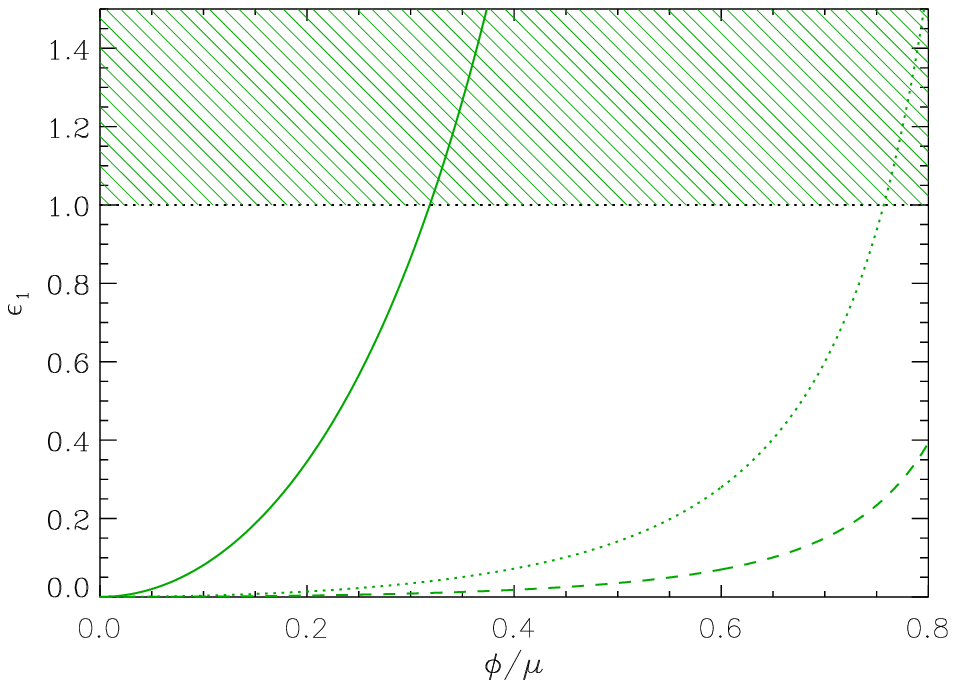}
\caption{The small field potential $\ln \Vall$ on the left and the
  first Hubble flow function $\epsone$ on the right. Inflation occurs
  for $\epsone<1$, for small values of the field. The three curves
  correspond respectively to $p=2$, $p=4$ and $p=6$ from the left to
  the right (from Ref.~\cite{Martin:2006rs}).}
\label{fig:sfpot}
\end{center}
\end{figure}

Small field inflation can be described by the action
(\ref{eq:actionsingle}) when the potential reads
\begin{equation}
\label{eq:sfpot}
\Vmat(\mat) = M^4 \left[1 - \left(\dfrac{\mat}{\mu} \right)^p \right].
\end{equation}
The inflation model parameters are the energy scale $M$, the power $p$
and the vacuum expectation value scale $\mu$. As can be seen in
Fig.~\ref{fig:sfpot}, inflation proceeds for small initial field
values and stops when $\epsone(\mat_\uend)=1$. Notice that the
potential in Eq.~(\ref{eq:sfpot}) is negative for $\mat>\mu$ which
means that the above description is no longer correct. This is however
not an issue since it occurs well after the end of inflation and the
basic effects of the reheating are already encoded in our extra
parameter $R_\urad$.

\subsubsection{Observable parameters}

Since flatness is inherited from inflation, the $\Lambda$CDM
cosmological model is described by the density parameters associated
with the different cosmological fluids: $\OmegaB$, $\OmegaCDM$ plus
the Hubble parameter today $\hubble_\zero$. The cosmological constant
is fixed by $\OmegaL = 1-\OmegaB - \OmegaCDM$. However, in order to
minimise the parameter degeneracies with respect to the CMB angular
power spectra, it is more convenient to perform the MCMC sampling on
the equivalent set of parameters $\OmegaB h^2$, $\OmegaCDM h^2$, the
optical depth $\optdepth$ and the quantity $\theta$ which measures the
ratio of the sound horizon to the angular diameter distance ($h$ is
the reduced Hubble parameter today)~\cite{Lewis:2002ah}.

Similarly for the potential parameters, as can be seen in
Eqs.~(\ref{eq:srpowerscal}) and (\ref{eq:srpowertens}), the overall
amplitude of the CMB anisotropies is proportional to the Hubble
parameter squared and thus to the potential $\Vall$ [see
Eq.~(\ref{eq:hubblesquare})]. Since the amplitude of the cosmological
perturbations is a well measured quantity, the data may be more
efficiently used by directly sampling the primordial amplitude of the
scalar power spectrum $\calP_*=\calP_\zeta(\kwav_*)$ instead of $M$
($\kwav_*$ being a fixed observable wavenumber: $\kwav_*/a_\zero=0.05
\Mpc^{-1}$).

However, the numerical method used to integrate the perturbations
during inflation requires the input of a numerical value for $M$ to
predict the value of $\calP_*$. In fact, one can use the trick
described in Ref.~\cite{Martin:2006rs}: under a rescaling $\Vall
\rightarrow s \Vall$, the power spectrum scales as $\calP_\zeta(\kwav)
\rightarrow s \calP_\zeta(s^{1/2} \kwav)$ at fixed $R_\urad$. The idea
is therefore to integrate the perturbations with an artificial
normalisation of the potential, for instance $M=1$, and then
analytically rescale $M$ from unity to its physical value that would
be associated with $\calP_*$. The required value of $s$ is given by
the ratio $\calP_*/\calPnum_\diamond$, where $\calPnum_\diamond$ is the
amplitude of the scalar power spectrum stemming from the numerical
integration with $M=1$ and evaluated at $\kwav_\diamond = \kwav_*
s^{-1/2}$. Still, it is not really straightforward to determine $s$
since both $\calP_\zeta$ and $\kwav$ change simultaneously. The last
subtlety is to remark that $\kwav_\diamond = \kwav_*$ if instead of
considering $R_\urad$ fixed, one considers the rescaling of $M$ at
fixed $\lnR$, with $\lnR$ defined by
\begin{equation}
\label{eq:defR}
\lnR \equiv \ln R_{\urad}+\dfrac{1}{4} \ln\left(\kappa ^4\rho
_{\uend}\right).
\end{equation}
The parameters $\reheat$ and $R_\urad$ differ only by $\rho_\uend$,
the energy density at the end of inflation which is uniquely
determined from $M$, $\mu$ and $p$. It will be therefore more
convenient to sample the model parameters $\calP_*$ and $\lnR$
(together with $\mu$ and $p$) rather than $M$ and $R_\urad$.

\subsubsection{Priors}

The prior probability distributions for the base cosmological
parameters $\OmegaB h^2$, $\OmegaCDM h^2$, $\tau$ and $\theta$ have
been chosen as wide top hat uniform distribution centred over their
current preferred value~\cite{Spergel:2006hy,Lewis:2002ah}.

Concerning the inflaton potential parameters, their priors can be
chosen according to various theoretical
prejudices~\cite{Martin:2006rs}. Since $\calP_*$ is related to the
energy scale during inflation, a uniform prior has been considered
around a value compatible with the amplitude of the cosmological
perturbations: $\ln(10^{10}\calP_*) \in [2.7,4.0]$. For the scale
$\mu$ associated with the vacuum expectation value of $\mat$, we have
considered two priors. The first includes smaller and larger values
than the Planck mass: $\kappa \mu \in[1/10,10]$. The second prior is
also uniform but include values much larger than the Planck mass:
$\kappa \mu \in [1/10,100]$. Finally, a uniform prior is chosen for
the power $p$ on $[2.4,10]$ ($p=2$ is a particular case, see
Ref.~\cite{Martin:2006rs}).

It remains to express our prior knowledge on the reheating parameter
$\lnR$. As previously mentioned, we are assuming that gravity can
still be described classically which only makes sense if the energy
densities involved remain smaller than the Planck energy scale, namely
for $\kappa^4 \rho_\uend < 1$. On the other side of the energy
spectrum, the success of big-bang nucleosynthesis (BBN) requires that
the universe is radiation dominated at that time, thus $\rho_\ureh >
\rho_\unuc$ with $\rho_\unuc \simeq 1\MeV^4$ (and $\rho_\uend >
\rho_\ureh$). Moreover, we will assume that during reheating the
expansion of the universe can be described as dominated by a
cosmological fluid of pressure $P$ and energy density $\rho$. In this
case, in order to satisfy the strong and dominant energy conditions in
General Relativity, one has $-1/3<P/\rho<1$ (notice that $P/\rho \le
-1/3$ would be inflation). From Eqs.~(\ref{eq:defRrad}) and
(\ref{eq:defR}), the resulting bounds read
\begin{equation}
\label{eq:lnRbounds}
\frac{1}{4} \ln \left(\kappa ^4 \rho _{\unuc}\right) < \lnR <
-\frac{1}{12} \ln \left(\kappa ^4 \rho _{\unuc}\right) +\frac13 \ln
\left(\kappa ^4 \rho _{\uend}\right),
\end{equation}
and an uniform prior on $\lnR$ have been chosen in between.

\subsubsection{Results}

The data sets used to constrain the small field $\Lambda$CDM model are
the WMAP third year data together with the Hubble Space Telescope
(HST) measurements ($\hubble_\zero = 72 \pm 8\,
\mathrm{km/s/Mpc}$~\cite{Freedman:2000cf}) and a top hat prior on the
age of the universe between $10\,\mathrm{Gyrs}$ and $20
\,\mathrm{Gyrs}$.
\begin{figure}
\begin{center}
\includegraphics[width=10cm]{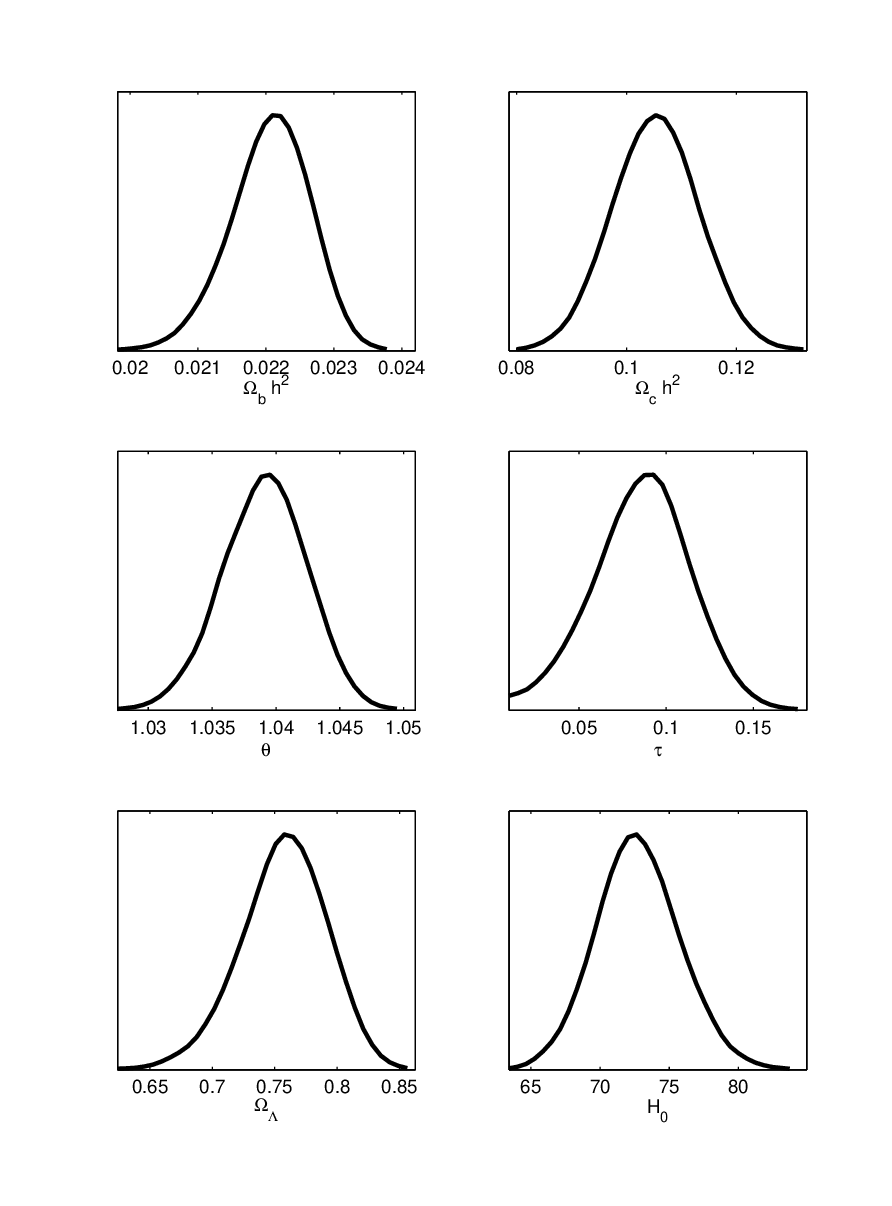}
\caption{Posterior probability distributions of the base cosmological
  parameters in the small field $\Lambda$CDM inflation model.}
\label{fig:sfcosmo}
\end{center}
\end{figure}
The resulting marginalised posterior probability distributions for the
base cosmological parameters are represented in
Fig.~\ref{fig:sfcosmo}. They are not significantly affected by the
various prior choices on $\mu$ and their corresponding mean values and
confidence intervals are compatible with the current state or
art~\cite{Spergel:2006hy}.

\begin{figure}
\begin{center}
\includegraphics[width=8cm]{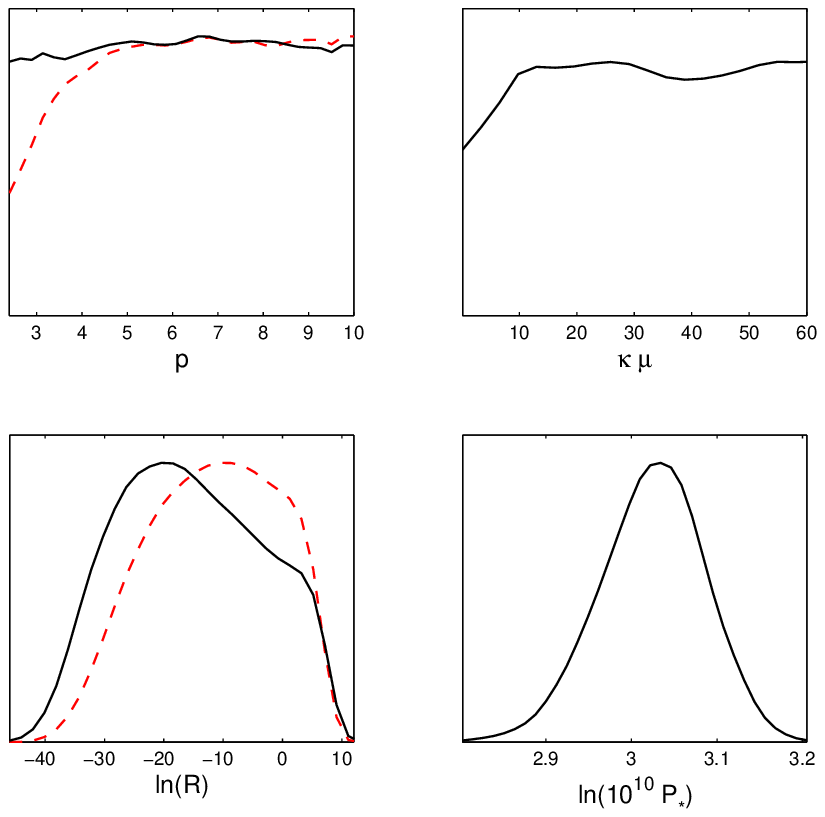}
\caption{Marginalised probability distributions for the small field
  inflation parameters. The results coming from the prior $\kappa \mu
  \in [0.1,100]$ are represented by solid black lines whereas, when
  they differ, the posteriors associated with the $\kappa \mu$ prior
  in $[0.1,10]$ are plotted as dashed red lines~\cite{Martin:2006rs}.}
\label{fig:sfinf}
\end{center}
\end{figure}

The constraints obtained on the small field inflation parameters are
showed in Fig.~\ref{fig:sfinf}. As expected, the allowed range for the
power spectra amplitude $\calP_*$ is narrow. Concerning the above
panels their interpretation require some precautions. Indeed, the
probability that $p$ takes small values depends on our theoretical
prejudice on how big the field expectation value of the inflaton may
be. If $\mu$ is allowed to be much greater than $\mpl$ then all $p$
are equiprobable. On the other hand, if $\kappa \mu$ cannot take
values bigger than $10$ then small field inflation models with
$p\simeq 2$ are disfavoured. The $\mu$ posterior shows on its own
that, independently of the $p$ values, $\kappa \mu > 10$ is slightly
preferred by the data. Since these posteriors are marginalised over
the other parameters, the previous statements are necessarily robust
with respect to any reheating model, in the framework of our
modelisation.

But more than being a nuisance parameter, Fig.~\ref{fig:sfinf} shows
that $\lnR$ is also mildly constrained by the data: the probability
distribution of $\ln R$ has a lower bound slightly above the prior
$\ln R > -46$ given by Eq.~(\ref{eq:lnRbounds}). Although this is not
obvious, the upper cut-off seen in the $\lnR$ posterior comes from the
upper bound of Eq.~(\ref{eq:lnRbounds}) (see Ref.~\cite{Martin:2006rs}
for a more detailed discussion). For $\kappa\mu$ in $[0.1,100]$, we
finally obtain at $95\%$ confidence level\footnote{The dependence of
the $\lnR$ posterior distribution with respect to the $\kappa \mu$
prior disappears as soon as $\kappa\mu$ is allowed to be greater than
$10$.}
\begin{equation}
\label{eq:lnRmin}
\ln R > -34.
\end{equation}

Plugging this inequality into Eq.~(\ref{eq:defR}) constrains some
properties of the reheating era. This result can be understood by
looking at Fig.~\ref{fig:scales}. Since varying the reheating
properties allow the observable window to move along the inflaton
potential, it is not surprising that some part of the potential may be
preferred from a data point of view. For the small field models, a
more involved analysis would show that the spectral index of the
scalar power spectrum gets away from what is allowed by the data when
$\lnR$ becomes too small~\cite{Martin:2006rs}. This is precisely why
the current WMAP data lead to the bound (\ref{eq:lnRmin}).

\section{Conclusion}

As a conclusion, we would like to discuss the future directions
associated with the possibility of constraining some basic properties
of the reheating era with the CMB data.

In the small field $\Lambda$CDM inflation analysed in the previous
section, the bound found in Eq.~(\ref{eq:lnRmin}) can be further
explored by being more specific on the way the universe reheated. As a
toy example, if one assumes that reheating proceeded with a constant
equation of state $P=\wstate \rho$, then Eq.~(\ref{eq:defR})
simplifies into
\begin{equation}
\label{eq:lnRw}
\lnR = \dfrac{1-3\wstate}{12+12\wstate} \ln\left(\kappa^4
\rho_\ureh\right) + \dfrac{1+3\wstate}{6+6\wstate} \ln \left(\kappa^4
\rho_\uend\right).
\end{equation}
\begin{figure}
\begin{center}
\includegraphics[height=9cm]{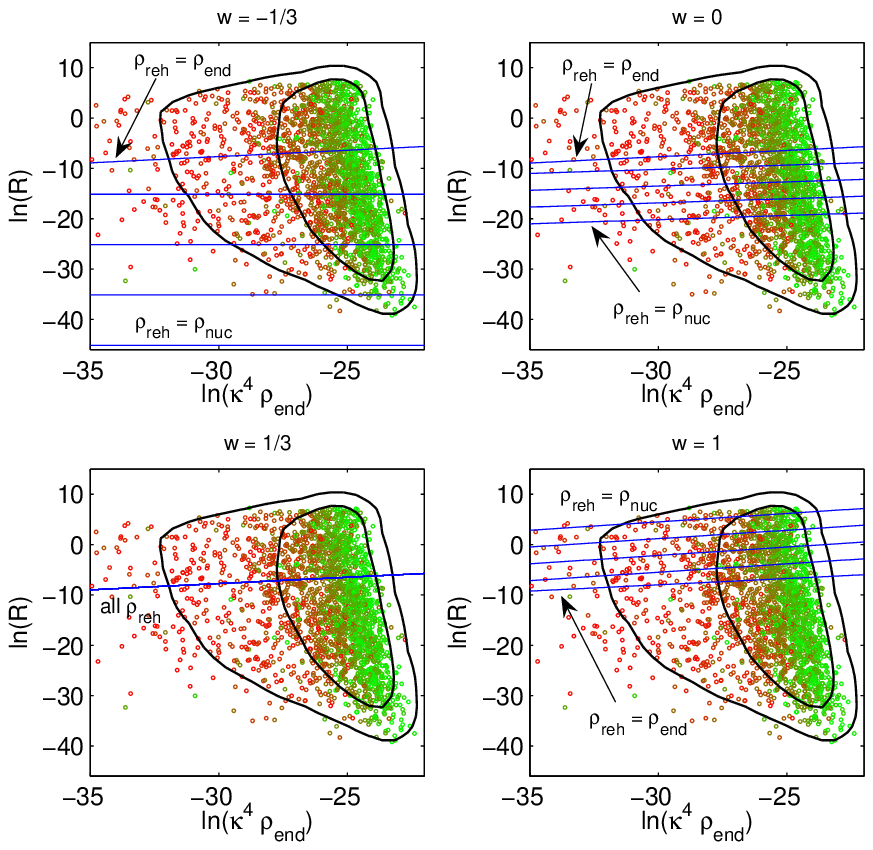}
\caption{One and two-sigma confidence intervals (solid contours) of
  the two-dimensional marginalised posteriors (point density) in the
  plane $[\ln R, \ln (\kappa ^4 \rho _{\uend})]$ for the small field
  models. The parameter $\kappa\mu$ varies in $[0.1,100]$ from red to
  green. The four panels correspond to the situation in which the
  universe reheated with a constant equation of state $P=\wstate
  \rho$. In each panel, the solid lines correspond to different values
  of the reheating temperature $(1/4) \ln(\kappa^4 \rho_{\ureh})$
  ranging from $-45$, $-35$, $-25$, $-15$ to $-(1/4)\ln(\kappa^4
  \rho_{\uend})$.}
\label{fig:wreh}
\end{center}
\end{figure}

In the parameter plane $[\lnR,\ln\left(\kappa^4 \rho_\uend\right)]$,
for a given value of the reheating energy, Eq.~(\ref{eq:lnRw})
corresponds to a straight line. In Fig.~\ref{fig:wreh}, five of these
lines exploring the range $\rho_\unuc<\rho_\ureh<\rho_\uend$ have been
superimposed to the two-dimensional probability distributions
associated with the small field model, and this for four different
equations of state having $\wstate \gtrsim -1/3, \wstate=0,
\wstate=1/3$ and $\wstate \lesssim 1$, respectively.

At can be seen on the top left frame, for $\wstate \gtrsim -1/3$, low
values of the reheating temperature are out of the confidence
contours. From a more robust analysis using importance sampling, one
would find in that case that, at $95\%$ of confidence, $\rho_\ureh >
2\TeV$~\cite{Martin:2006rs}. Of course this bound is not really
impressive and close to the limits already set by BBN, moreover, it
holds only for small field models and a quite extreme equation of
state. However, is shows that it is already possible to get some
information on the reheating era in a given model of inflation from
the CMB data only. This is precisely on this point that particle
physics models may be decisive. Indeed, once the inflaton couplings to
the other particles are specified, the properties of the reheating era
are fixed and a given particle physics model would appear as one curve
parametrised by some coupling constants in the plane $[\ln R, \ln
(\kappa ^4 \rho _{\uend})]$ of Fig.~\ref{fig:wreh}. With the incoming
flow of more accurate cosmological data, this may be an interesting
way of constraining inflation as well as particle physics at very high
energy.

\section*{Acknowledgments}

I would like to thank J\'er\^ome Martin for his comments on the
manuscript as well as the organisers of the ``Inflation+25''
conference for the outstanding scientific atmosphere that took place
during this colloquium. This work was partially supported by the
Belgian Federal Office for Scientific, Technical and Cultural Affairs
through the Inter-University Attraction Pole grant P6/11.

\bibliographystyle{JHEP} \bibliography{colloquium}

\providecommand{\href}[2]{#2}\begingroup\raggedright\begin{thebibliography}{10}

\bibitem{Spergel:2006hy}
D.~N. Spergel {\em et.~al.}, {\it Wilkinson microwave anisotropy probe (wmap)
  three year results: Implications for cosmology},
  \href{http://xxx.lanl.gov/abs/astro-ph/0603449}{{\tt astro-ph/0603449}}.

\bibitem{Bassett:2005xm}
B.~A. Bassett, S.~Tsujikawa, and D.~Wands, {\it Inflation dynamics and
  reheating},  {\em Rev. Mod. Phys.} {\bf 78} (2006) 537--589,
  [\href{http://xxx.lanl.gov/abs/astro-ph/0507632}{{\tt astro-ph/0507632}}].

\bibitem{Allahverdi:2006iq}
R.~Allahverdi, K.~Enqvist, J.~Garcia-Bellido, and A.~Mazumdar, {\it Gauge
  invariant mssm inflaton},  {\em Phys. Rev. Lett.} {\bf 97} (2006) 191304,
  [\href{http://xxx.lanl.gov/abs/hep-ph/0605035}{{\tt hep-ph/0605035}}].

\bibitem{Allahverdi:2006we}
R.~Allahverdi, K.~Enqvist, J.~Garcia-Bellido, A.~Jokinen, and A.~Mazumdar, {\it
  Mssm flat direction inflation: Slow roll, stability, fine tunning and
  reheating},  \href{http://xxx.lanl.gov/abs/hep-ph/0610134}{{\tt
  hep-ph/0610134}}.

\bibitem{Tye:2006uv}
S.~H.~H. Tye, {\it Brane inflation: String theory viewed from the cosmos},
  \href{http://xxx.lanl.gov/abs/hep-th/0610221}{{\tt hep-th/0610221}}.

\bibitem{Schwarz:2001vv}
D.~J. Schwarz, C.~A. Terrero-Escalante, and A.~A. Garcia, {\it Higher order
  corrections to primordial spectra from cosmological inflation},  {\em Phys.
  Lett.} {\bf B517} (2001) 243--249,
  [\href{http://xxx.lanl.gov/abs/astro-ph/0106020}{{\tt astro-ph/0106020}}].

\bibitem{Lyth:1998xn}
D.~H. Lyth and A.~Riotto, {\it Particle physics models of inflation and the
  cosmological density perturbation},  {\em Phys. Rept.} {\bf 314} (1999)
  1--146, [\href{http://xxx.lanl.gov/abs/hep-ph/9807278}{{\tt
  hep-ph/9807278}}].

\bibitem{Mukhanov:1990me}
V.~F. Mukhanov, H.~A. Feldman, and R.~H. Brandenberger, {\it Theory of
  cosmological perturbations. part 1. classical perturbations. part 2. quantum
  theory of perturbations. part 3. extensions},  {\em Phys. Rept.} {\bf 215}
  (1992) 203--333.

\bibitem{Stewart:1993bc}
E.~D. Stewart and D.~H. Lyth, {\it A more accurate analytic calculation of the
  spectrum of cosmological perturbations produced during inflation},  {\em
  Phys. Lett.} {\bf B302} (1993) 171--175,
  [\href{http://xxx.lanl.gov/abs/gr-qc/9302019}{{\tt gr-qc/9302019}}].

\bibitem{Martin:1999wa}
J.~Martin and D.~J. Schwarz, {\it The precision of slow-roll predictions for
  the cmbr anisotropies},  {\em Phys. Rev.} {\bf D62} (2000) 103520,
  [\href{http://xxx.lanl.gov/abs/astro-ph/9911225}{{\tt astro-ph/9911225}}].

\bibitem{Martin:2006rs}
J.~Martin and C.~Ringeval, {\it Inflation after wmap3: Confronting the
  slow-roll and exact power spectra with cmb data},  {\em JCAP} {\bf 0608}
  (2006) 009, [\href{http://xxx.lanl.gov/abs/astro-ph/0605367}{{\tt
  astro-ph/0605367}}].

\bibitem{Liddle:1994dx}
A.~R. Liddle, P.~Parsons, and J.~D. Barrow, {\it Formalizing the slow roll
  approximation in inflation},  {\em Phys. Rev.} {\bf D50} (1994) 7222--7232,
  [\href{http://xxx.lanl.gov/abs/astro-ph/9408015}{{\tt astro-ph/9408015}}].

\bibitem{Martin:1997zd}
J.~Martin and D.~J. Schwarz, {\it The influence of cosmological transitions on
  the evolution of density perturbations},  {\em Phys. Rev.} {\bf D57} (1998)
  3302--3316, [\href{http://xxx.lanl.gov/abs/gr-qc/9704049}{{\tt
  gr-qc/9704049}}].

\bibitem{Martin:2004um}
J.~Martin, {\it Inflationary cosmological perturbations of quantum- mechanical
  origin},  {\em Lect. Notes Phys.} {\bf 669} (2005) 199--244,
  [\href{http://xxx.lanl.gov/abs/hep-th/0406011}{{\tt hep-th/0406011}}].

\bibitem{Schwarz:2004tz}
D.~J. Schwarz and C.~A. Terrero-Escalante, {\it Primordial fluctuations and
  cosmological inflation after wmap 1.0},  {\em JCAP} {\bf 0408} (2004) 003,
  [\href{http://xxx.lanl.gov/abs/hep-ph/0403129}{{\tt hep-ph/0403129}}].

\bibitem{deOliveira:2005mf}
H.~P. de~Oliveira and C.~A. Terrero-Escalante, {\it Troubles for observing the
  inflaton potential},  {\em JCAP} {\bf 0601} (2006) 024,
  [\href{http://xxx.lanl.gov/abs/astro-ph/0511660}{{\tt astro-ph/0511660}}].

\bibitem{Felder:2000hj}
G.~N. Felder {\em et.~al.}, {\it Dynamics of symmetry breaking and tachyonic
  preheating},  {\em Phys. Rev. Lett.} {\bf 87} (2001) 011601,
  [\href{http://xxx.lanl.gov/abs/hep-ph/0012142}{{\tt hep-ph/0012142}}].

\bibitem{Kofman:1997yn}
L.~Kofman, A.~D. Linde, and A.~A. Starobinsky, {\it Towards the theory of
  reheating after inflation},  {\em Phys. Rev.} {\bf D56} (1997) 3258--3295,
  [\href{http://xxx.lanl.gov/abs/hep-ph/9704452}{{\tt hep-ph/9704452}}].

\bibitem{Garcia-Bellido:1997wm}
J.~Garcia-Bellido and A.~D. Linde, {\it Preheating in hybrid inflation},  {\em
  Phys. Rev.} {\bf D57} (1998) 6075--6088,
  [\href{http://xxx.lanl.gov/abs/hep-ph/9711360}{{\tt hep-ph/9711360}}].

\bibitem{Senoguz:2004vu}
V.~N. Senoguz and Q.~Shafi, {\it Reheat temperature in supersymmetric hybrid
  inflation models},  {\em Phys. Rev.} {\bf D71} (2005) 043514,
  [\href{http://xxx.lanl.gov/abs/hep-ph/0412102}{{\tt hep-ph/0412102}}].

\bibitem{Podolsky:2005bw}
D.~I. Podolsky, G.~N. Felder, L.~Kofman, and M.~Peloso, {\it Equation of state
  and beginning of thermalization after preheating},  {\em Phys. Rev.} {\bf
  D73} (2006) 023501, [\href{http://xxx.lanl.gov/abs/hep-ph/0507096}{{\tt
  hep-ph/0507096}}].

\bibitem{Desroche:2005yt}
M.~Desroche, G.~N. Felder, J.~M. Kratochvil, and A.~Linde, {\it Preheating in
  new inflation},  {\em Phys. Rev.} {\bf D71} (2005) 103516,
  [\href{http://xxx.lanl.gov/abs/hep-th/0501080}{{\tt hep-th/0501080}}].

\bibitem{Allahverdi:2006wh}
R.~Allahverdi and A.~Mazumdar, {\it Towards a successful reheating within
  supersymmetry},  \href{http://xxx.lanl.gov/abs/hep-ph/0603244}{{\tt
  hep-ph/0603244}}.

\bibitem{Liddle:2003as}
A.~R. Liddle and S.~M. Leach, {\it How long before the end of inflation were
  observable perturbations produced?},  {\em Phys. Rev.} {\bf D68} (2003)
  103503, [\href{http://xxx.lanl.gov/abs/astro-ph/0305263}{{\tt
  astro-ph/0305263}}].

\bibitem{Barriga:2000nk}
J.~Barriga, E.~Gaztanaga, M.~G. Santos, and S.~Sarkar, {\it On the apm power
  spectrum and the cmb anisotropy: Evidence for a phase transition during
  inflation?},  {\em Mon. Not. Roy. Astron. Soc.} {\bf 324} (2001) 977,
  [\href{http://xxx.lanl.gov/abs/astro-ph/0011398}{{\tt astro-ph/0011398}}].

\bibitem{Martin:2003sg}
J.~Martin and C.~Ringeval, {\it Superimposed oscillations in the wmap data?},
  {\em Phys. Rev.} {\bf D69} (2004) 083515,
  [\href{http://xxx.lanl.gov/abs/astro-ph/0310382}{{\tt astro-ph/0310382}}].

\bibitem{Martin:2004iv}
J.~Martin and C.~Ringeval, {\it Addendum to ``superimposed oscillations in the
  wmap data?''},  {\em Phys. Rev.} {\bf D69} (2004) 127303,
  [\href{http://xxx.lanl.gov/abs/astro-ph/0402609}{{\tt astro-ph/0402609}}].

\bibitem{Martin:2004yi}
J.~Martin and C.~Ringeval, {\it Exploring the superimposed oscillations
  parameter space},  {\em JCAP} {\bf 0501} (2005) 007,
  [\href{http://xxx.lanl.gov/abs/hep-ph/0405249}{{\tt hep-ph/0405249}}].

\bibitem{Easther:2004vq}
R.~Easther, W.~H. Kinney, and H.~Peiris, {\it Observing trans-planckian
  signatures in the cosmic microwave background},  {\em JCAP} {\bf 0505} (2005)
  009, [\href{http://xxx.lanl.gov/abs/astro-ph/0412613}{{\tt
  astro-ph/0412613}}].

\bibitem{Hunt:2004vt}
P.~Hunt and S.~Sarkar, {\it Multiple inflation and the wmap 'glitches'},  {\em
  Phys. Rev.} {\bf D70} (2004) 103518,
  [\href{http://xxx.lanl.gov/abs/astro-ph/0408138}{{\tt astro-ph/0408138}}].

\bibitem{Covi:2006ci}
L.~Covi, J.~Hamann, A.~Melchiorri, A.~Slosar, and I.~Sorbera, {\it Inflation
  and wmap three year data: Features have a future!},  {\em Phys. Rev.} {\bf
  D74} (2006) 083509, [\href{http://xxx.lanl.gov/abs/astro-ph/0606452}{{\tt
  astro-ph/0606452}}].

\bibitem{Martin:2002vn}
J.~Martin and D.~J. Schwarz, {\it Wkb approximation for inflationary
  cosmological perturbations},  {\em Phys. Rev.} {\bf D67} (2003) 083512,
  [\href{http://xxx.lanl.gov/abs/astro-ph/0210090}{{\tt astro-ph/0210090}}].

\bibitem{Casadio:2004ru}
R.~Casadio, F.~Finelli, M.~Luzzi, and G.~Venturi, {\it Improved wkb analysis of
  cosmological perturbations},  {\em Phys. Rev.} {\bf D71} (2005) 043517,
  [\href{http://xxx.lanl.gov/abs/gr-qc/0410092}{{\tt gr-qc/0410092}}].

\bibitem{Casadio:2005xv}
R.~Casadio, F.~Finelli, M.~Luzzi, and G.~Venturi, {\it Higher order slow-roll
  predictions for inflation},  {\em Phys. Lett.} {\bf B625} (2005) 1--6,
  [\href{http://xxx.lanl.gov/abs/gr-qc/0506043}{{\tt gr-qc/0506043}}].

\bibitem{Casadio:2006wb}
R.~Casadio, F.~Finelli, A.~Kamenshchik, M.~Luzzi, and G.~Venturi, {\it Method
  of comparison equations for cosmological perturbations},  {\em JCAP} {\bf
  0604} (2006) 011, [\href{http://xxx.lanl.gov/abs/gr-qc/0603026}{{\tt
  gr-qc/0603026}}].

\bibitem{Gordon:2000hv}
C.~Gordon, D.~Wands, B.~A. Bassett, and R.~Maartens, {\it Adiabatic and entropy
  perturbations from inflation},  {\em Phys. Rev.} {\bf D63} (2001) 023506,
  [\href{http://xxx.lanl.gov/abs/astro-ph/0009131}{{\tt astro-ph/0009131}}].

\bibitem{Noh:2001ia}
H.~Noh and J.-c. Hwang, {\it Inflationary spectra in generalized gravity:
  Unified forms},  {\em Phys. Lett.} {\bf B515} (2001) 231--237,
  [\href{http://xxx.lanl.gov/abs/astro-ph/0107069}{{\tt astro-ph/0107069}}].

\bibitem{DiMarco:2002eb}
F.~Di~Marco, F.~Finelli, and R.~Brandenberger, {\it Adiabatic and isocurvature
  perturbations for multifield generalized einstein models},  {\em Phys. Rev.}
  {\bf D67} (2003) 063512,
  [\href{http://xxx.lanl.gov/abs/astro-ph/0211276}{{\tt astro-ph/0211276}}].

\bibitem{DiMarco:2005nq}
F.~Di~Marco and F.~Finelli, {\it Slow-roll inflation for generalized two-field
  lagrangians},  {\em Phys. Rev.} {\bf D71} (2005) 123502,
  [\href{http://xxx.lanl.gov/abs/astro-ph/0505198}{{\tt astro-ph/0505198}}].

\bibitem{Salopek:1988qh}
D.~S. Salopek, J.~R. Bond, and J.~M. Bardeen, {\it Designing density
  fluctuation spectra in inflation},  {\em Phys. Rev.} {\bf D40} (1989) 1753.

\bibitem{Grivell:1999wc}
I.~J. Grivell and A.~R. Liddle, {\it Inflaton potential reconstruction without
  slow-roll},  {\em Phys. Rev.} {\bf D61} (2000) 081301,
  [\href{http://xxx.lanl.gov/abs/astro-ph/9906327}{{\tt astro-ph/9906327}}].

\bibitem{Adams:2001vc}
J.~A. Adams, B.~Cresswell, and R.~Easther, {\it Inflationary perturbations from
  a potential with a step},  {\em Phys. Rev.} {\bf D64} (2001) 123514,
  [\href{http://xxx.lanl.gov/abs/astro-ph/0102236}{{\tt astro-ph/0102236}}].

\bibitem{Tsujikawa:2002qx}
S.~Tsujikawa, D.~Parkinson, and B.~A. Bassett, {\it Correlation-consistency
  cartography of the double inflation landscape},  {\em Phys. Rev.} {\bf D67}
  (2003) 083516, [\href{http://xxx.lanl.gov/abs/astro-ph/0210322}{{\tt
  astro-ph/0210322}}].

\bibitem{Parkinson:2004yx}
D.~Parkinson, S.~Tsujikawa, B.~A. Bassett, and L.~Amendola, {\it Testing for
  double inflation with wmap},  {\em Phys. Rev.} {\bf D71} (2005) 063524,
  [\href{http://xxx.lanl.gov/abs/astro-ph/0409071}{{\tt astro-ph/0409071}}].

\bibitem{Makarov:2005uh}
A.~Makarov, {\it On the accuracy of slow-roll inflation given current
  observational constraints},  {\em Phys. Rev.} {\bf D72} (2005) 083517,
  [\href{http://xxx.lanl.gov/abs/astro-ph/0506326}{{\tt astro-ph/0506326}}].

\bibitem{Chen:2006xj}
X.~Chen, R.~Easther, and E.~A. Lim, {\it Large non-gaussianities in single
  field inflation},  \href{http://xxx.lanl.gov/abs/astro-ph/0611645}{{\tt
  astro-ph/0611645}}.

\bibitem{Lewis:1999bs}
A.~Lewis, A.~Challinor, and A.~Lasenby, {\it Efficient computation of cmb
  anisotropies in closed frw models},  {\em Astrophys. J.} {\bf 538} (2000)
  473--476, [\href{http://xxx.lanl.gov/abs/astro-ph/9911177}{{\tt
  astro-ph/9911177}}].

\bibitem{Lewis:2002ah}
A.~Lewis and S.~Bridle, {\it Cosmological parameters from cmb and other data: a
  monte- carlo approach},  {\em Phys. Rev.} {\bf D66} (2002) 103511,
  [\href{http://xxx.lanl.gov/abs/astro-ph/0205436}{{\tt astro-ph/0205436}}].

\bibitem{Peiris:2006sj}
H.~Peiris and R.~Easther, {\it Slow roll reconstruction: Constraints on
  inflation from the 3 year wmap dataset},  {\em JCAP} {\bf 0610} (2006) 017,
  [\href{http://xxx.lanl.gov/abs/astro-ph/0609003}{{\tt astro-ph/0609003}}].

\bibitem{Peiris:2006ug}
H.~Peiris and R.~Easther, {\it Recovering the inflationary potential and
  primordial power spectrum with a slow roll prior},
  \href{http://xxx.lanl.gov/abs/astro-ph/0603587}{{\tt astro-ph/0603587}}.

\bibitem{Kinney:2006qm}
W.~H. Kinney, E.~W. Kolb, A.~Melchiorri, and A.~Riotto, {\it Inflation model
  constraints from the wilkinson microwave anisotropy probe three-year data},
  {\em Phys. Rev.} {\bf D74} (2006) 023502,
  [\href{http://xxx.lanl.gov/abs/astro-ph/0605338}{{\tt astro-ph/0605338}}].

\bibitem{Brax:2004xh}
P.~Brax, C.~van~de Bruck, and A.-C. Davis, {\it Brane world cosmology},  {\em
  Rept. Prog. Phys.} {\bf 67} (2004) 2183--2232,
  [\href{http://xxx.lanl.gov/abs/hep-th/0404011}{{\tt hep-th/0404011}}].

\bibitem{Damour:1992we}
T.~Damour and G.~Esposito-Farese, {\it Tensor multiscalar theories of
  gravitation},  {\em Class. Quant. Grav.} {\bf 9} (1992) 2093--2176.

\bibitem{Damour:1993id}
T.~Damour and K.~Nordtvedt, {\it Tensor - scalar cosmological models and their
  relaxation toward general relativity},  {\em Phys. Rev.} {\bf D48} (1993)
  3436--3450.

\bibitem{Koshelev:2005wk}
N.~A. Koshelev, {\it Adiabatic and entropy perturbations in inflationary models
  based on non-linear sigma model},  {\em Grav. Cosmol.} {\bf 10} (2004)
  289--294, [\href{http://xxx.lanl.gov/abs/astro-ph/0501600}{{\tt
  astro-ph/0501600}}].

\bibitem{Ringeval:2005yn}
C.~Ringeval, P.~Brax, v.~de~Bruck, Carsten, and A.-C. Davis, {\it Boundary
  inflation and the wmap data},  {\em Phys. Rev.} {\bf D73} (2006) 064035,
  [\href{http://xxx.lanl.gov/abs/astro-ph/0509727}{{\tt astro-ph/0509727}}].

\bibitem{Langlois:2002bb}
D.~Langlois, {\it Brane cosmology: An introduction},  {\em Prog. Theor. Phys.
  Suppl.} {\bf 148} (2003) 181--212,
  [\href{http://xxx.lanl.gov/abs/hep-th/0209261}{{\tt hep-th/0209261}}].

\bibitem{Maartens:2003tw}
R.~Maartens, {\it Brane-world gravity},  {\em Living Rev. Rel.} {\bf 7} (2004)
  7, [\href{http://xxx.lanl.gov/abs/gr-qc/0312059}{{\tt gr-qc/0312059}}].

\bibitem{Schimd:2004nq}
C.~Schimd, J.-P. Uzan, and A.~Riazuelo, {\it Weak lensing in scalar-tensor
  theories of gravity},  {\em Phys. Rev.} {\bf D71} (2005) 083512,
  [\href{http://xxx.lanl.gov/abs/astro-ph/0412120}{{\tt astro-ph/0412120}}].

\bibitem{Lukas:1998tt}
A.~Lukas, B.~A. Ovrut, K.~S. Stelle, and D.~Waldram, {\it Heterotic m-theory in
  five dimensions},  {\em Nucl. Phys.} {\bf B552} (1999) 246--290,
  [\href{http://xxx.lanl.gov/abs/hep-th/9806051}{{\tt hep-th/9806051}}].

\bibitem{Lukas:1999yn}
A.~Lukas, B.~A. Ovrut, and D.~Waldram, {\it Boundary inflation},  {\em Phys.
  Rev.} {\bf D61} (2000) 023506,
  [\href{http://xxx.lanl.gov/abs/hep-th/9902071}{{\tt hep-th/9902071}}].

\bibitem{Brax:2000xk}
P.~Brax and A.~C. Davis, {\it Cosmological solutions of supergravity in
  singular spaces},  {\em Phys. Lett.} {\bf B497} (2001) 289--295,
  [\href{http://xxx.lanl.gov/abs/hep-th/0011045}{{\tt hep-th/0011045}}].

\bibitem{Kobayashi:2002pw}
S.~Kobayashi and K.~Koyama, {\it Cosmology with radion and bulk scalar field in
  two branes model},  {\em JHEP} {\bf 12} (2002) 056,
  [\href{http://xxx.lanl.gov/abs/hep-th/0210029}{{\tt hep-th/0210029}}].

\bibitem{Esposito-Farese:2000ij}
G.~Esposito-Farese and D.~Polarski, {\it Scalar-tensor gravity in an
  accelerating universe},  {\em Phys. Rev.} {\bf D63} (2001) 063504,
  [\href{http://xxx.lanl.gov/abs/gr-qc/0009034}{{\tt gr-qc/0009034}}].

\bibitem{Martin:2005bp}
J.~Martin, C.~Schimd, and J.-P. Uzan, {\it Testing for w < -1 in the solar
  system},  {\em Phys. Rev. Lett.} {\bf 96} (2006) 061303,
  [\href{http://xxx.lanl.gov/abs/astro-ph/0510208}{{\tt astro-ph/0510208}}].

\bibitem{Carter:1997pb}
B.~Carter, {\it Brane dynamics for treatment of cosmic strings and vortons},
  \href{http://xxx.lanl.gov/abs/hep-th/9705172}{{\tt hep-th/9705172}}.

\bibitem{Liddle:1993fq}
A.~R. Liddle and D.~H. Lyth, {\it The cold dark matter density perturbation},
  {\em Phys. Rept.} {\bf 231} (1993) 1--105,
  [\href{http://xxx.lanl.gov/abs/astro-ph/9303019}{{\tt astro-ph/9303019}}].

\bibitem{Turner:1983he}
M.~S. Turner, {\it Coherent scalar field oscillations in an expanding
  universe},  {\em Phys. Rev.} {\bf D28} (1983) 1243.

\bibitem{Niemeyer:2002kh}
J.~C. Niemeyer, R.~Parentani, and D.~Campo, {\it Minimal modifications of the
  primordial power spectrum from an adiabatic short distance cutoff},  {\em
  Phys. Rev.} {\bf D66} (2002) 083510,
  [\href{http://xxx.lanl.gov/abs/hep-th/0206149}{{\tt hep-th/0206149}}].

\bibitem{Weinberg:2004kf}
S.~Weinberg, {\it Must cosmological perturbations remain non-adiabatic after
  multi-field inflation?},  {\em Phys. Rev.} {\bf D70} (2004) 083522,
  [\href{http://xxx.lanl.gov/abs/astro-ph/0405397}{{\tt astro-ph/0405397}}].

\bibitem{Lemoine:2006sc}
M.~Lemoine and J.~Martin, {\it Neutralino dark matter and the curvaton},
  \href{http://xxx.lanl.gov/abs/astro-ph/0611948}{{\tt astro-ph/0611948}}.

\bibitem{Page:2006hz}
L.~Page {\em et.~al.}, {\it Three year wilkinson microwave anisotropy probe
  (wmap) observations: Polarization analysis},
  \href{http://xxx.lanl.gov/abs/astro-ph/0603450}{{\tt astro-ph/0603450}}.

\bibitem{Hinshaw:2006ia}
G.~Hinshaw {\em et.~al.}, {\it Three-year wilkinson microwave anisotropy probe
  (wmap) observations: Temperature analysis},
  \href{http://xxx.lanl.gov/abs/astro-ph/0603451}{{\tt astro-ph/0603451}}.

\bibitem{Jarosik:2006ib}
N.~Jarosik {\em et.~al.}, {\it Three-year wilkinson microwave anisotropy probe
  (wmap) observations: Beam profiles, data processing, radiometer
  characterization and systematic error limits},
  \href{http://xxx.lanl.gov/abs/astro-ph/0603452}{{\tt astro-ph/0603452}}.

\bibitem{Freedman:2000cf}
W.~L. Freedman {\em et.~al.}, {\it Final results from the hubble space
  telescope key project to measure the hubble constant},  {\em Astrophys. J.}
  {\bf 553} (2001) 47--72,
  [\href{http://xxx.lanl.gov/abs/astro-ph/0012376}{{\tt astro-ph/0012376}}].

\end{thebibliography}\endgroup
\end{document}